\documentclass[conference,a4paper]{IEEEtran}
\usepackage{cite}
\usepackage{amsmath,amssymb,amsfonts,bm}
\usepackage[linesnumbered,ruled,vlined]{algorithm2e}
\usepackage{amsthm}
\usepackage{dsfont}
\usepackage{algorithmic}
\usepackage{subfigure}
\usepackage{graphicx}
\usepackage{textcomp}
 \usepackage{threeparttable}
\usepackage{xcolor}
\usepackage{bbm}
\usepackage[nolist,withpage]{acronym}
\usepackage{array}
\usepackage{tikz}
\usepackage{pgfplots}
\pgfplotsset{compat=newest}

\definecolor{copperrose}{rgb}{0.6, 0.4, 0.4}
\definecolor{azure}{rgb}{0.0, 0.5, 1.0}
\definecolor{ashgrey}{rgb}{0.7, 0.75, 0.71}
\definecolor{chestnut}{rgb}{0.8, 0.36, 0.36}
\definecolor{airforceblue}{rgb}{0.36, 0.54, 0.66}
\definecolor{cadmiumorange}{rgb}{0.93, 0.53, 0.18}
\definecolor{bleudefrance}{rgb}{0.19, 0.55, 0.91}
\definecolor{carolinablue}{rgb}{0.6, 0.73, 0.89}
\definecolor{blue(ncs)}{rgb}{0.0, 0.53, 0.74}
\definecolor{dodgerblue}{rgb}{0.12, 0.56, 1.0}
\definecolor{cssgreen}{rgb}{0.0, 0.5, 0.0}
\definecolor{cadmiumgreen}{rgb}{0.0, 0.42, 0.24}
\definecolor{cadmiumorange}{rgb}{0.93, 0.53, 0.18}
\definecolor{amaranth}{rgb}{0.9, 0.17, 0.31}
\definecolor{bluegray}{rgb}{0.4, 0.6, 0.8}
\definecolor{cerulean}{rgb}{0.0, 0.48, 0.65}
\definecolor{ceil}{rgb}{0.57, 0.63, 0.81}

\usepackage{etoolbox}
\makeatletter
\newif\if@in@acrolist
\AtBeginEnvironment{acronym}{\@in@acrolisttrue}
\newrobustcmd{\LU}[2]{\if@in@acrolist#1\else#2\fi}

\newcommand{\ACF}[1]{{\@in@acrolisttrue\acf{#1}}}

\begin{document}

%\chapter*{Acronyms}
%\renewcommand{\titulonome}{Acronyms}%
%\renewcommand{\prepbynome}{UFC.33 Team}%

%\begin{singlespace}
\begin{acronym}[LTE-Advanced]%\addtolength{\itemsep}{-0.5\baselineskip}
  \acro{2G}{Second Generation}
  \acro{3-DAP}{3-Dimensional Assignment Problem}
  \acro{3G}{3$^\text{rd}$~Generation}
  \acro{3GPP}{3$^\text{rd}$~Generation Partnership Project}
  \acro{4G}{4$^\text{th}$~Generation}
  \acro{5G}{5$^\text{th}$~Generation}
  \acro{AA}{Antenna Array}
  \acro{AC}{Admission Control}
  \acro{AD}{Attack-Decay}
  \acro{ADC}{analog-to-digital converter}
  \acro{ADMM}{alternating direction method of multipliers}
  \acro{ADSL}{Asymmetric Digital Subscriber Line}
  \acro{AHW}{Alternate Hop-and-Wait}
  \acro{AirComp}{over-the-air computation}
  \acro{AMC}{Adaptive Modulation and Coding}
  \acro{ANN}{artificial neural network}
  \acro{AP}{\LU{A}{a}ccess \LU{P}{p}oint}
  \acro{APA}{Adaptive Power Allocation}
  \acro{ARMA}{Autoregressive Moving Average}
  \acro{ARQ}{\LU{A}{a}utomatic \LU{R}{r}epeat \LU{R}{r}equest}
  \acro{ATES}{Adaptive Throughput-based Efficiency-Satisfaction Trade-Off}
  \acro{AWGN}{additive white Gaussian noise}
  \acro{BAA}{\LU{B}{b}roadband \LU{A}{a}nalog \LU{A}{a}ggregation}
  \acro{BB}{Branch and Bound}
  \acro{BCD}{block coordinate descent}
  \acro{BD}{Block Diagonalization}
  \acro{BER}{Bit Error Rate}
  \acro{BF}{Best Fit}
  \acro{BFD}{bidirectional full duplex}
  \acro{BLER}{BLock Error Rate}
  \acro{BPC}{Binary Power Control}
  \acro{BPSK}{Binary Phase-Shift Keying}
  \acro{BRA}{Balanced Random Allocation}
  \acro{BS}{base station}
  \acro{BSUM}{block successive upper-bound minimization}
  \acro{CAP}{Combinatorial Allocation Problem}
  \acro{CAPEX}{Capital Expenditure}
  \acro{CBF}{Coordinated Beamforming}
  \acro{CBR}{Constant Bit Rate}
  \acro{CBS}{Class Based Scheduling}
  \acro{CC}{Congestion Control}
  \acro{CDF}{Cumulative Distribution Function}
  \acro{CDMA}{Code-Division Multiple Access}
  \acro{CE}{\LU{C}{c}hannel \LU{E}{e}stimation}
  \acro{CL}{Closed Loop}
  \acro{CLPC}{Closed Loop Power Control}
  \acro{CML}{centralized machine learning}
  \acro{CNR}{Channel-to-Noise Ratio}
  \acro{CNN}{\LU{C}{c}onvolutional \LU{N}{n}eural \LU{N}{n}etwork}
  \acro{CP}{computation point}
  \acro{CPA}{Cellular Protection Algorithm}
  \acro{CPICH}{Common Pilot Channel}
  \acro{CoCoA}{\LU{C}{c}ommunication efficient distributed dual \LU{C}{c}oordinate \LU{A}{a}scent}
  \acro{CoMAC}{\LU{C}{c}omputation over \LU{M}{m}ultiple-\LU{A}{a}ccess \LU{C}{c}hannels}
  \acro{CoMP}{Coordinated Multi-Point}
  \acro{CQI}{Channel Quality Indicator}
  \acro{CRM}{Constrained Rate Maximization}
	\acro{CRN}{Cognitive Radio Network}
  \acro{CS}{Coordinated Scheduling}
  \acro{CSI}{\LU{C}{c}hannel \LU{S}{s}tate \LU{I}{i}nformation}
  \acro{CSMA}{\LU{C}{c}arrier \LU{S}{s}ense \LU{M}{m}ultiple \LU{A}{a}ccess}
  \acro{CUE}{Cellular User Equipment}
  \acro{D2D}{device-to-device}
  \acro{DAC}{digital-to-analog converter}
  \acro{DC}{direct current}
  \acro{DCA}{Dynamic Channel Allocation}
  \acro{DE}{Differential Evolution}
  \acro{DFT}{Discrete Fourier Transform}
%  \acro{DIST}{Distance-based Grouping}
  \acro{DIST}{Distance}
  \acro{DL}{downlink}
  \acro{DMA}{Double Moving Average}
  \acro{DML}{Distributed Machine Learning}
  \acro{DMRS}{demodulation reference signal}
  \acro{D2DM}{D2D Mode}
  \acro{DMS}{D2D Mode Selection}
  \acro{DPC}{Dirty Paper Coding}
  \acro{DRA}{Dynamic Resource Assignment}
  \acro{DSA}{Dynamic Spectrum Access}
  \acro{DSGD}{\LU{D}{d}istributed \LU{S}{s}tochastic \LU{G}{g}radient \LU{D}{d}escent}
  \acro{DSM}{Delay-based Satisfaction Maximization}
  \acro{ECC}{Electronic Communications Committee}
  \acro{EFLC}{Error Feedback Based Load Control}
  \acro{EI}{Efficiency Indicator}
  \acro{eNB}{Evolved Node B}
  \acro{EPA}{Equal Power Allocation}
  \acro{EPC}{Evolved Packet Core}
  \acro{EPS}{Evolved Packet System}
  \acro{E-UTRAN}{Evolved Universal Terrestrial Radio Access Network}
  \acro{ES}{Exhaustive Search}
  %\acro{FD}{full duplex}
  \acro{FC}{\LU{F}{f}usion \LU{C}{c}enter}
  \acro{FD}{\LU{F}{f}ederated \LU{D}{d}istillation}
  \acro{FDD}{frequency divisionov duplex}
  \acro{FDM}{Frequency Division Multiplexing}
  \acro{FDMA}{\LU{F}{f}requency \LU{D}{d}ivision \LU{M}{m}ultiple \LU{A}{a}ccess}
  \acro{FedAvg}{\LU{F}{f}ederated \LU{A}{a}veraging}
  \acro{FER}{Frame Erasure Rate}
  \acro{FF}{Fast Fading}
  \acro{FL}{federated learning}
  \acro{FSB}{Fixed Switched Beamforming}
  \acro{FST}{Fixed SNR Target}
  \acro{FTP}{File Transfer Protocol}
  \acro{GA}{Genetic Algorithm}
  \acro{GBR}{Guaranteed Bit Rate}
  \acro{GLR}{Gain to Leakage Ratio}
  \acro{GOS}{Generated Orthogonal Sequence}
  \acro{GPL}{GNU General Public License}
  \acro{GRP}{Grouping}
  \acro{HARQ}{Hybrid Automatic Repeat Request}
  \acro{HD}{half-duplex}
  \acro{HMS}{Harmonic Mode Selection}
  \acro{HOL}{Head Of Line}
  \acro{HSDPA}{High-Speed Downlink Packet Access}
  \acro{HSPA}{High Speed Packet Access}
  \acro{HTTP}{HyperText Transfer Protocol}
  \acro{ICMP}{Internet Control Message Protocol}
  \acro{ICI}{Intercell Interference}
  \acro{ID}{Identification}
  \acro{IETF}{Internet Engineering Task Force}
  \acro{ILP}{Integer Linear Program}
  \acro{JRAPAP}{Joint RB Assignment and Power Allocation Problem}
  \acro{UID}{Unique Identification}
  \acro{IID}{\LU{I}{i}ndependent and \LU{I}{i}dentically \LU{D}{d}istributed}
  \acro{IIR}{Infinite Impulse Response}
  \acro{ILP}{Integer Linear Problem}
  \acro{IMT}{International Mobile Telecommunications}
  \acro{INV}{Inverted Norm-based Grouping}
  \acro{IoT}{Internet of Things}
%  \acro{IP}{Internet Protocol}
  \acro{IP}{Integer Programming}
  \acro{IPv6}{Internet Protocol Version 6}
  \acro{ISD}{Inter-Site Distance}
  \acro{ISI}{Inter Symbol Interference}
  \acro{ITU}{International Telecommunication Union}
  \acro{JAFM}{joint assignment and fairness maximization}
  \acro{JAFMA}{joint assignment and fairness maximization algorithm}
  \acro{JOAS}{Joint Opportunistic Assignment and Scheduling}
  \acro{JOS}{Joint Opportunistic Scheduling}
  \acro{JP}{Joint Processing}
	\acro{JS}{Jump-Stay}
  \acro{KKT}{Karush-Kuhn-Tucker}
  \acro{L3}{Layer-3}
  \acro{LAC}{Link Admission Control}
  \acro{LA}{Link Adaptation}
  \acro{LC}{Load Control}
  \acro{LDC}{\LU{L}{l}earning-\LU{D}{d}riven \LU{C}{c}ommunication}
  \acro{LOS}{line of sight}
  \acro{LP}{Linear Programming}
  \acro{LTE}{Long Term Evolution}
	\acro{LTE-A}{\ac{LTE}-Advanced}
  \acro{LTE-Advanced}{Long Term Evolution Advanced}
  \acro{M2M}{Machine-to-Machine}
  \acro{MAC}{multiple access channel}
  \acro{MANET}{Mobile Ad hoc Network}
  \acro{MC}{Modular Clock}
  \acro{MCS}{Modulation and Coding Scheme}
  \acro{MDB}{Measured Delay Based}
  \acro{MDI}{Minimum D2D Interference}
  \acro{MF}{Matched Filter}
  \acro{MG}{Maximum Gain}
  \acro{MH}{Multi-Hop}
  \acro{MIMO}{\LU{M}{m}ultiple \LU{I}{i}nput \LU{M}{m}ultiple \LU{O}{o}utput}
  \acro{MINLP}{mixed integer nonlinear programming}
  \acro{MIP}{mixed integer programming}
  \acro{MISO}{multiple input single output}
  \acro{ML}{machine learning}
  \acro{MLE}{maximum likelihood estimator}
  \acro{MLWDF}{Modified Largest Weighted Delay First}
  \acro{MME}{Mobility Management Entity}
  \acro{MMSE}{minimum mean squared error}
  \acro{MOS}{Mean Opinion Score}
  \acro{MPF}{Multicarrier Proportional Fair}
  \acro{MRA}{Maximum Rate Allocation}
  \acro{MR}{Maximum Rate}
  \acro{MRC}{Maximum Ratio Combining}
  \acro{MRT}{Maximum Ratio Transmission}
  \acro{MRUS}{Maximum Rate with User Satisfaction}
  \acro{MS}{Mode Selection}
  \acro{MSE}{\LU{M}{m}ean \LU{S}{s}quared \LU{E}{e}rror}
  \acro{MSI}{Multi-Stream Interference}
  \acro{MTC}{Machine-Type Communication}
  \acro{MTSI}{Multimedia Telephony Services over IMS}
  \acro{MTSM}{Modified Throughput-based Satisfaction Maximization}
  \acro{MU-MIMO}{Multi-User Multiple Input Multiple Output}
  \acro{MU}{Multi-User}
  \acro{NAS}{Non-Access Stratum}
  \acro{NB}{Node B}
	\acro{NCL}{Neighbor Cell List}
  \acro{NLP}{Nonlinear Programming}
  \acro{NLOS}{non-line of sight}
  \acro{NMSE}{normalized mean square error}
  \acro{NOMA}{\LU{N}{n}on-\LU{O}{o}rthogonal \LU{M}{m}ultiple \LU{A}{a}ccess}
  \acro{NORM}{Normalized Projection-based Grouping}
  \acro{NP}{non-polynomial time}
  \acro{NRT}{Non-Real Time}
  \acro{NSPS}{National Security and Public Safety Services}
  \acro{O2I}{Outdoor to Indoor}
  \acro{OFDMA}{\LU{O}{o}rthogonal \LU{F}{f}requency \LU{D}{d}ivision \LU{M}{m}ultiple \LU{A}{a}ccess}
  \acro{OFDM}{Orthogonal Frequency Division Multiplexing}
  \acro{OFPC}{Open Loop with Fractional Path Loss Compensation}
	\acro{O2I}{Outdoor-to-Indoor}
  \acro{OL}{Open Loop}
  \acro{OLPC}{Open-Loop Power Control}
  \acro{OL-PC}{Open-Loop Power Control}
  \acro{OPEX}{Operational Expenditure}
  \acro{ORB}{Orthogonal Random Beamforming}
  \acro{JO-PF}{Joint Opportunistic Proportional Fair}
  \acro{OSI}{Open Systems Interconnection}
  \acro{PAIR}{D2D Pair Gain-based Grouping}
  \acro{PAPR}{Peak-to-Average Power Ratio}
  \acro{P2P}{Peer-to-Peer}
  \acro{PC}{Power Control}
  \acro{PCI}{Physical Cell ID}
  \acro{PDCCH}{physical downlink control channel}
  \acro{PDD}{penalty dual decomposition}
  \acro{PDF}{Probability Density Function}
  \acro{PER}{Packet Error Rate}
  \acro{PF}{Proportional Fair}
  \acro{P-GW}{Packet Data Network Gateway}
  \acro{PL}{Pathloss}
  \acro{RLT}{reformulation linearization technique}
  \acro{PRB}{Physical Resource Block}
  \acro{PROJ}{Projection-based Grouping}
  \acro{ProSe}{Proximity Services}
%  \acro{PS}{Packet Scheduling}
%  \acro{PS}{phase shifter}
  \acro{PS}{\LU{P}{p}arameter \LU{S}{s}erver}
  \acro{PSO}{Particle Swarm Optimization}
  \acro{PUCCH}{physical uplink control channel}
  \acro{PZF}{Projected Zero-Forcing}
  \acro{QAM}{quadrature amplitude modulation}
  \acro{QoS}{quality of service}
  \acro{QPSK}{quadrature phase shift keying}
  \acro{QCQP}{quadratically constrained quadratic programming}
  \acro{RAISES}{Reallocation-based Assignment for Improved Spectral Efficiency and Satisfaction}
  \acro{RAN}{Radio Access Network}
  \acro{RA}{Resource Allocation}
  \acro{RAT}{Radio Access Technology}
  \acro{RATE}{Rate-based}
  \acro{RB}{resource block}
  \acro{RBG}{Resource Block Group}
  \acro{REF}{Reference Grouping}
  \acro{ReLU}{rectified linear unit}
  \acro{RF}{radio frequency}
  \acro{RLC}{Radio Link Control}
  \acro{RM}{Rate Maximization}
  \acro{RNC}{Radio Network Controller}
  \acro{RND}{Random Grouping}
  \acro{RRA}{Radio Resource Allocation}
  \acro{RRM}{\LU{R}{r}adio \LU{R}{r}esource \LU{M}{m}anagement}
  \acro{RSCP}{Received Signal Code Power}
  \acro{RSRP}{reference signal receive power}
  \acro{RSRQ}{Reference Signal Receive Quality}
  \acro{RR}{Round Robin}
  \acro{RRC}{Radio Resource Control}
  \acro{RSSI}{received signal strength indicator}
  \acro{RT}{Real Time}
  \acro{RU}{Resource Unit}
  \acro{RUNE}{RUdimentary Network Emulator}
  \acro{RV}{Random Variable}
  \acro{SAA}{Small Argument Approximation}
  \acro{SAC}{Session Admission Control}
  \acro{SCM}{Spatial Channel Model}
  \acro{SC-FDMA}{Single Carrier - Frequency Division Multiple Access}
  \acro{SD}{Soft Dropping}
  \acro{S-D}{Source-Destination}
  \acro{SDPC}{Soft Dropping Power Control}
  \acro{SDMA}{Space-Division Multiple Access}
  \acro{SDR}{semidefinite relaxation}
  \acro{SDP}{semidefinite programming}
  \acro{SER}{Symbol Error Rate}
  \acro{SES}{Simple Exponential Smoothing}
  \acro{S-GW}{Serving Gateway}
  \acro{SGD}{\LU{S}{s}tochastic \LU{G}{g}radient \LU{D}{d}escent}  
  \acro{SINR}{signal-to-interference-plus-noise ratio}
%   \acro{SI}{Satisfaction Indicator}
  \acro{SI}{self-interference}
  \acro{SIP}{Session Initiation Protocol}
  \acro{SISO}{\LU{S}{s}ingle \LU{I}{i}nput \LU{S}{s}ingle \LU{O}{o}utput}
  \acro{SIMO}{Single Input Multiple Output}
  \acro{SIR}{Signal to Interference Ratio}
  \acro{SLNR}{Signal-to-Leakage-plus-Noise Ratio}
  \acro{SMA}{Simple Moving Average}
  \acro{SNR}{\LU{S}{s}ignal-to-\LU{N}{n}oise \LU{R}{r}atio}
  \acro{SORA}{Satisfaction Oriented Resource Allocation}
  \acro{SORA-NRT}{Satisfaction-Oriented Resource Allocation for Non-Real Time Services}
  \acro{SORA-RT}{Satisfaction-Oriented Resource Allocation for Real Time Services}
  \acro{SPF}{Single-Carrier Proportional Fair}
  \acro{SRA}{Sequential Removal Algorithm}
  \acro{SRS}{sounding reference signal}
  \acro{SU-MIMO}{Single-User Multiple Input Multiple Output}
  \acro{SU}{Single-User}
  \acro{SVD}{Singular Value Decomposition}
  \acro{SVM}{\LU{S}{s}upport \LU{V}{v}ector \LU{M}{m}achine}
  \acro{TCP}{Transmission Control Protocol}
  \acro{TDD}{time division duplex}
  \acro{TDMA}{\LU{T}{t}ime \LU{D}{d}ivision \LU{M}{m}ultiple \LU{A}{a}ccess}
  \acro{TNFD}{three node full duplex}
  \acro{TETRA}{Terrestrial Trunked Radio}
  \acro{TP}{Transmit Power}
  \acro{TPC}{Transmit Power Control}
  \acro{TTI}{transmission time interval}
  \acro{TTR}{Time-To-Rendezvous}
  \acro{TSM}{Throughput-based Satisfaction Maximization}
  \acro{TU}{Typical Urban}
  \acro{UE}{\LU{U}{u}ser \LU{E}{e}quipment}
  \acro{UEPS}{Urgency and Efficiency-based Packet Scheduling}
  \acro{UL}{uplink}
  \acro{UMTS}{Universal Mobile Telecommunications System}
  \acro{URI}{Uniform Resource Identifier}
  \acro{URM}{Unconstrained Rate Maximization}
  \acro{VR}{Virtual Resource}
  \acro{VoIP}{Voice over IP}
  \acro{WAN}{Wireless Access Network}
  \acro{WCDMA}{Wideband Code Division Multiple Access}
  \acro{WF}{Water-filling}
  \acro{WiMAX}{Worldwide Interoperability for Microwave Access}
  \acro{WINNER}{Wireless World Initiative New Radio}
  \acro{WLAN}{Wireless Local Area Network}
  \acro{WMMSE}{weighted minimum mean square error}
  \acro{WMPF}{Weighted Multicarrier Proportional Fair}
  \acro{WPF}{Weighted Proportional Fair}
  \acro{WSN}{Wireless Sensor Network}
  \acro{WWW}{World Wide Web}
  \acro{XIXO}{(Single or Multiple) Input (Single or Multiple) Output}
  \acro{ZF}{Zero-Forcing}
  \acro{ZMCSCG}{Zero Mean Circularly Symmetric Complex Gaussian}
\end{acronym}
%\end{singlespace}

\title{A Novel Channel Coding Scheme \\ for Digital Multiple Access Computing
}

\author{Xiaojing Yan, Saeed Razavikia, Carlo Fischione\\
	\normalsize School of Electrical Engineering and Computer Science, KTH Royal Institute of Technology, Stockholm, Sweden\\
	\normalsize Email: \{xiay, sraz, carlofi\}@kth.se
}

\newtheorem{theorem}{Theorem}
\newtheorem{prop}{Proposition}
\newtheorem{lem}{Lemma}

\maketitle
\thispagestyle{empty}
\pagestyle{empty}

\begin{abstract}
In this paper, we consider the ChannelComp framework, which facilitates the computation of desired functions by multiple transmitters over a common receiver using digital modulations across a multiple access channel. While ChannelComp currently offers a broad framework for computation by designing digital constellations for over-the-air computation and employing symbol-level encoding, encoding the repeated transmissions of the same symbol and using the corresponding received sequence may significantly improve the computation performance and reduce the encoding complexity. In this paper, we propose an enhancement involving the encoding of the repetitive transmission of the same symbol at each transmitter over multiple time slots and the design of constellation diagrams, with the aim of minimizing computational errors. We frame this enhancement as an optimization problem, which jointly identifies the constellation diagram and the channel code for repetition, which we call ReChCompCode. To manage the computational complexity of the optimization, we divide it into two tractable subproblems. Through numerical experiments, we evaluate the performance of ReChCompCode. The simulation results reveal that ReChCompCode can reduce the computation error by approximately up to $30$~dB compared to standard ChannelComp, particularly for product functions. 
\end{abstract}


\section{Introduction}

\acresetall

The rise of \ac{IoT}-centric applications increases the demand for reliable and spectrum-efficient wireless protocols to gather data from distributed smart devices. A notable strategy to enhance spectral efficiency is \ac{AirComp}~\cite{csahin2023survey}. \ac{AirComp} is an analog communication scheme that concurrently allocates all devices over the same frequency resources and leverages the inherent waveform superposition of wireless signals. \ac{AirComp} has shown its effectiveness in supporting machine learning services~\cite{hellstrom2022wireless,razavikia2023blind}. Specifically, integrating federated learning with AirComp turns out to be promising in resource-constrained settings, often yielding superior model outcomes due to diverse device knowledge.

Nonetheless, the dependency of \ac{AirComp} on analog modulations raises questions about its suit for present-day \ac{IoT} applications~\cite{goldenbaum2013harnessing}. The analog communication lacks an effective noise countermeasure, merely diminishing the estimation error, which compromises its reliability~\cite{hellstrom2022wireless}. Addressing these analog communication challenges, a digital computation framework named ChannelComp has been introduced in~\cite{razavikia2023computing}. This method, which constitutes a paradigm shift compared to \ac{AirComp}, is a framework for general function computation over \ac{MAC} using digital modulations~\cite{saeed2023ChannelComp,razavikia2023sumcomp}.

Standard digital communications rely on channel coding to ensure robust communication, embedding redundant information to compensate channel noise~\cite{shannon1948mathematical}. In the \ac{AirComp}'s literature, the primary methodology centers on nested lattice codes' linearity~\cite{nazer2007computation}. This enables computations over Gaussian channels. A nested lattice coding-based computation scheme further reinforces function computation against noise~\cite{goldenbaum2014nomographic}. Yet, such codes, including the coded \ac{AirComp} tailored for federated learning model aggregation \cite{zhang2022coded}, are constrained to simple digital modulations, such as ASK, and do not generalize in terms of modulation formats or function computations.

In this paper, we propose repetition of the transmissions and an associated channel coding scheme for the ChannelComp framework to provide a reliable and spectral-efficient communication strategy. To reduce the computation error, we design both the digital constellation and the coding of the repeated symbol sequence, where we call the resulting code ReChCompCode. Specifically, we propose the conditions for a valid function computation, which leads to an optimization problem to determine the constellation diagram of digital modulation over the time slots. The proposed optimization problem involves the parameters of both channel and source codes. To handle the complexity, we decompose the problem into two subproblems. The first subproblem becomes a \ac{SDP} problem by relaxing non-convex constraints. The second subproblem turns into an \ac{MIP} format \cite{wolsey2007mixed}, manageable with the branch and bound method \cite{pedroso2011optimization}. Finally, we assess the performance of ReChCompCode for providing a reliable computation through numerical experiments.

% ========================

% \tikzset{every picture/.style={line width=0.75pt}}

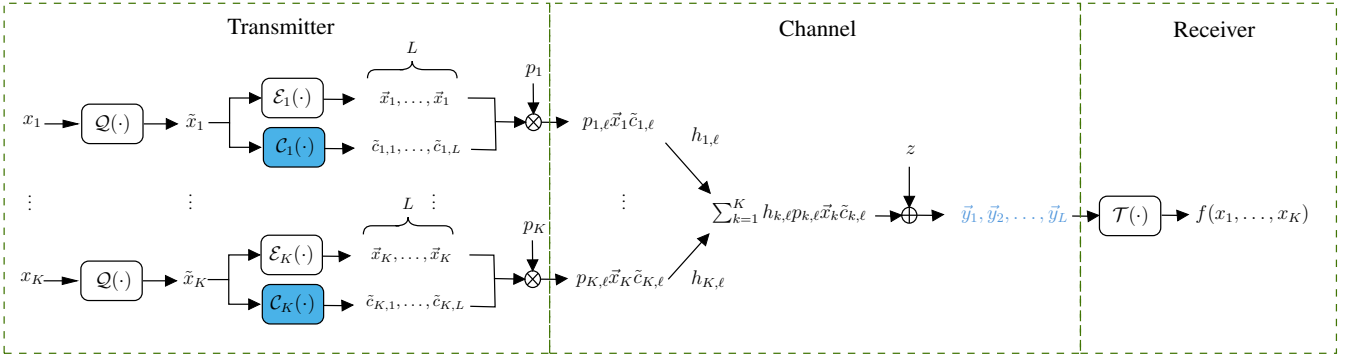
\begin{figure*}[!t]
\centering
\scalebox{0.62}{

\tikzset{every picture/.style={line width=0.75pt}} %set default line width to 0.75pt        

\begin{tikzpicture}[x=0.75pt,y=0.75pt,yscale=-1,xscale=1]

%Straight Lines [id:da9872405594483162] 
\draw    (55,107) -- (80,107) ;
\draw [shift={(82,107)}, rotate = 180] [fill={rgb, 255:red, 0; green, 0; blue, 0 }  ][line width=0.08]  [draw opacity=0] (12,-3) -- (0,0) -- (12,3) -- cycle    ;
%Rounded Rect [id:dp7290842175896497] 
\draw   (85,97.4) .. controls (85,93.87) and (87.87,91) .. (91.4,91) -- (127.93,91) .. controls (131.47,91) and (134.33,93.87) .. (134.33,97.4) -- (134.33,116.6) .. controls (134.33,120.13) and (131.47,123) .. (127.93,123) -- (91.4,123) .. controls (87.87,123) and (85,120.13) .. (85,116.6) -- cycle ;
%Straight Lines [id:da6110587387132285] 
\draw    (135,107) -- (159,107) ;
\draw [shift={(162,107)}, rotate = 180] [fill={rgb, 255:red, 0; green, 0; blue, 0 }  ][line width=0.08]  [draw opacity=0] (8.93,-4.29) -- (0,0) -- (8.93,4.29) -- cycle    ;
%Straight Lines [id:da13230189815749793] 
\draw    (187,106) -- (204,106) -- (204,86) -- (226,86) ;
\draw [shift={(229,86)}, rotate = 180] [fill={rgb, 255:red, 0; green, 0; blue, 0 }  ][line width=0.08]  [draw opacity=0] (8.93,-4.29) -- (0,0) -- (8.93,4.29) -- cycle    ;
%Straight Lines [id:da17997651767913325] 
\draw    (204,106) -- (204,126) -- (226,126) ;
\draw [shift={(229,126)}, rotate = 180] [fill={rgb, 255:red, 0; green, 0; blue, 0 }  ][line width=0.08]  [draw opacity=0] (8.93,-4.29) -- (0,0) -- (8.93,4.29) -- cycle    ;
%Rounded Rect [id:dp9592048607620245] 
\draw   (230,76.4) .. controls (230,72.87) and (232.87,70) .. (236.4,70) -- (272.93,70) .. controls (276.47,70) and (279.33,72.87) .. (279.33,76.4) -- (279.33,95.6) .. controls (279.33,99.13) and (276.47,102) .. (272.93,102) -- (236.4,102) .. controls (232.87,102) and (230,99.13) .. (230,95.6) -- cycle ;
%Rounded Rect [id:dp37862756756735094] 
\draw  [fill={rgb, 255:red, 73; green, 178; blue, 232 }  ,fill opacity=1 ] (231.21,116.4) .. controls (231.21,112.87) and (234.07,110) .. (237.61,110) -- (274.14,110) .. controls (277.68,110) and (280.54,112.87) .. (280.54,116.4) -- (280.54,135.6) .. controls (280.54,139.13) and (277.68,142) .. (274.14,142) -- (237.61,142) .. controls (234.07,142) and (231.21,139.13) .. (231.21,135.6) -- cycle ;
%Straight Lines [id:da052009437788159474] 
\draw    (395.67,86) -- (417.33,86.17) -- (417.83,105.17) -- (439.33,105.61) ;
\draw [shift={(442.33,105.67)}, rotate = 181.17] [fill={rgb, 255:red, 0; green, 0; blue, 0 }  ][line width=0.08]  [draw opacity=0] (8.93,-4.29) -- (0,0) -- (8.93,4.29) -- cycle    ;
%Straight Lines [id:da08655424568157932] 
\draw    (398.33,127.67) -- (418.33,128.17) -- (417.83,105.17) ;
%Flowchart: Summing Junction [id:dp9827946955723679] 
\draw   (442.33,105.67) .. controls (442.33,101.94) and (445.13,98.92) .. (448.58,98.92) .. controls (452.04,98.92) and (454.83,101.94) .. (454.83,105.67) .. controls (454.83,109.39) and (452.04,112.42) .. (448.58,112.42) .. controls (445.13,112.42) and (442.33,109.39) .. (442.33,105.67) -- cycle ; \draw   (444.16,100.89) -- (453,110.44) ; \draw   (453,100.89) -- (444.16,110.44) ;
%Straight Lines [id:da48455433715364493] 
\draw    (448.33,74.17) -- (448.55,95.92) ;
\draw [shift={(448.58,98.92)}, rotate = 269.42] [fill={rgb, 255:red, 0; green, 0; blue, 0 }  ][line width=0.08]  [draw opacity=0] (8.93,-4.29) -- (0,0) -- (8.93,4.29) -- cycle    ;
%Straight Lines [id:da12284694494021098] 
\draw    (279.5,87) -- (293.33,87.12) -- (301.34,86.79) ;
\draw [shift={(304.33,86.67)}, rotate = 177.66] [fill={rgb, 255:red, 0; green, 0; blue, 0 }  ][line width=0.08]  [draw opacity=0] (8.93,-4.29) -- (0,0) -- (8.93,4.29) -- cycle    ;
%Straight Lines [id:da9311412539041846] 
\draw    (554.76,122.29) -- (588.78,161.03) ;
\draw [shift={(590.76,163.29)}, rotate = 228.72] [fill={rgb, 255:red, 0; green, 0; blue, 0 }  ][line width=0.08]  [draw opacity=0] (8.93,-4.29) -- (0,0) -- (8.93,4.29) -- cycle    ;
%Straight Lines [id:da8744560813324391] 
\draw    (280.17,127) -- (294,127.12) -- (301,126.8) ;
\draw [shift={(304,126.67)}, rotate = 177.43] [fill={rgb, 255:red, 0; green, 0; blue, 0 }  ][line width=0.08]  [draw opacity=0] (8.93,-4.29) -- (0,0) -- (8.93,4.29) -- cycle    ;
%Straight Lines [id:da3563282934740797] 
\draw    (454.83,105.67) -- (468.66,105.78) -- (473.28,105.38) ;
\draw [shift={(476.26,105.11)}, rotate = 174.95] [fill={rgb, 255:red, 0; green, 0; blue, 0 }  ][line width=0.08]  [draw opacity=0] (8.93,-4.29) -- (0,0) -- (8.93,4.29) -- cycle    ;
%Straight Lines [id:da07213244321486201] 
\draw    (723.5,181.67) -- (737.33,181.78) -- (741.94,181.38) ;
\draw [shift={(744.93,181.11)}, rotate = 174.95] [fill={rgb, 255:red, 0; green, 0; blue, 0 }  ][line width=0.08]  [draw opacity=0] (8.93,-4.29) -- (0,0) -- (8.93,4.29) -- cycle    ;
%Flowchart: Or [id:dp9512002279527669] 
\draw   (744.93,181.11) .. controls (744.93,177.57) and (747.64,174.69) .. (750.99,174.69) .. controls (754.34,174.69) and (757.06,177.57) .. (757.06,181.11) .. controls (757.06,184.65) and (754.34,187.53) .. (750.99,187.53) .. controls (747.64,187.53) and (744.93,184.65) .. (744.93,181.11) -- cycle ; \draw   (744.93,181.11) -- (757.06,181.11) ; \draw   (750.99,174.69) -- (750.99,187.53) ;
%Straight Lines [id:da8977842223480288] 
\draw    (751,141.78) -- (750.99,171.69) ;
\draw [shift={(750.99,174.69)}, rotate = 270.01] [fill={rgb, 255:red, 0; green, 0; blue, 0 }  ][line width=0.08]  [draw opacity=0] (8.93,-4.29) -- (0,0) -- (8.93,4.29) -- cycle    ;
%Straight Lines [id:da7765070941156642] 
\draw    (757.06,181.11) -- (770.89,181.23) -- (775.5,180.82) ;
\draw [shift={(778.49,180.56)}, rotate = 174.95] [fill={rgb, 255:red, 0; green, 0; blue, 0 }  ][line width=0.08]  [draw opacity=0] (8.93,-4.29) -- (0,0) -- (8.93,4.29) -- cycle    ;
%Rounded Rect [id:dp48263180777178705] 
\draw   (904,172.07) .. controls (904,168.53) and (906.87,165.67) .. (910.4,165.67) -- (946.93,165.67) .. controls (950.47,165.67) and (953.33,168.53) .. (953.33,172.07) -- (953.33,191.27) .. controls (953.33,194.8) and (950.47,197.67) .. (946.93,197.67) -- (910.4,197.67) .. controls (906.87,197.67) and (904,194.8) .. (904,191.27) -- cycle ;
%Straight Lines [id:da3355239746631302] 
\draw    (882.06,181.78) -- (895.89,181.89) -- (900.5,181.49) ;
\draw [shift={(903.49,181.22)}, rotate = 174.95] [fill={rgb, 255:red, 0; green, 0; blue, 0 }  ][line width=0.08]  [draw opacity=0] (8.93,-4.29) -- (0,0) -- (8.93,4.29) -- cycle    ;
%Straight Lines [id:da8568092433099779] 
\draw    (954.06,181.11) -- (967.89,181.23) -- (972.5,180.82) ;
\draw [shift={(975.49,180.56)}, rotate = 174.95] [fill={rgb, 255:red, 0; green, 0; blue, 0 }  ][line width=0.08]  [draw opacity=0] (8.93,-4.29) -- (0,0) -- (8.93,4.29) -- cycle    ;
%Straight Lines [id:da264205511902484] 
\draw    (54.33,234.33) -- (79.33,234.33) ;
\draw [shift={(81.33,234.33)}, rotate = 180] [fill={rgb, 255:red, 0; green, 0; blue, 0 }  ][line width=0.08]  [draw opacity=0] (12,-3) -- (0,0) -- (12,3) -- cycle    ;
%Rounded Rect [id:dp32747932806669033] 
\draw   (84.33,224.73) .. controls (84.33,221.2) and (87.2,218.33) .. (90.73,218.33) -- (127.27,218.33) .. controls (130.8,218.33) and (133.67,221.2) .. (133.67,224.73) -- (133.67,243.93) .. controls (133.67,247.47) and (130.8,250.33) .. (127.27,250.33) -- (90.73,250.33) .. controls (87.2,250.33) and (84.33,247.47) .. (84.33,243.93) -- cycle ;
%Straight Lines [id:da8914823477670122] 
\draw    (134.33,234.33) -- (158.33,234.33) ;
\draw [shift={(161.33,234.33)}, rotate = 180] [fill={rgb, 255:red, 0; green, 0; blue, 0 }  ][line width=0.08]  [draw opacity=0] (8.93,-4.29) -- (0,0) -- (8.93,4.29) -- cycle    ;
%Straight Lines [id:da5849294822258673] 
\draw    (186.33,233.33) -- (203.33,233.33) -- (203.33,213.33) -- (225.33,213.33) ;
\draw [shift={(228.33,213.33)}, rotate = 180] [fill={rgb, 255:red, 0; green, 0; blue, 0 }  ][line width=0.08]  [draw opacity=0] (8.93,-4.29) -- (0,0) -- (8.93,4.29) -- cycle    ;
%Straight Lines [id:da05801492900348881] 
\draw    (203.33,233.33) -- (203.33,253.33) -- (225.33,253.33) ;
\draw [shift={(228.33,253.33)}, rotate = 180] [fill={rgb, 255:red, 0; green, 0; blue, 0 }  ][line width=0.08]  [draw opacity=0] (8.93,-4.29) -- (0,0) -- (8.93,4.29) -- cycle    ;
%Rounded Rect [id:dp012454103837826969] 
\draw   (229.33,203.73) .. controls (229.33,200.2) and (232.2,197.33) .. (235.73,197.33) -- (272.27,197.33) .. controls (275.8,197.33) and (278.67,200.2) .. (278.67,203.73) -- (278.67,222.93) .. controls (278.67,226.47) and (275.8,229.33) .. (272.27,229.33) -- (235.73,229.33) .. controls (232.2,229.33) and (229.33,226.47) .. (229.33,222.93) -- cycle ;
%Rounded Rect [id:dp23841884028504512] 
\draw  [fill={rgb, 255:red, 73; green, 178; blue, 232 }  ,fill opacity=1 ] (230.33,243.73) .. controls (230.33,240.2) and (233.2,237.33) .. (236.73,237.33) -- (273.27,237.33) .. controls (276.8,237.33) and (279.67,240.2) .. (279.67,243.73) -- (279.67,262.93) .. controls (279.67,266.47) and (276.8,269.33) .. (273.27,269.33) -- (236.73,269.33) .. controls (233.2,269.33) and (230.33,266.47) .. (230.33,262.93) -- cycle ;
%Straight Lines [id:da7454436425692794] 
\draw    (397,213.33) -- (416.67,213.5) -- (417.17,232.5) -- (438.67,232.94) ;
\draw [shift={(441.67,233)}, rotate = 181.17] [fill={rgb, 255:red, 0; green, 0; blue, 0 }  ][line width=0.08]  [draw opacity=0] (8.93,-4.29) -- (0,0) -- (8.93,4.29) -- cycle    ;
%Straight Lines [id:da7219472206446143] 
\draw    (397.67,255) -- (417.67,255.5) -- (417.17,232.5) ;
%Flowchart: Summing Junction [id:dp08920563050697061] 
\draw   (441.67,233) .. controls (441.67,229.27) and (444.46,226.25) .. (447.92,226.25) .. controls (451.37,226.25) and (454.17,229.27) .. (454.17,233) .. controls (454.17,236.73) and (451.37,239.75) .. (447.92,239.75) .. controls (444.46,239.75) and (441.67,236.73) .. (441.67,233) -- cycle ; \draw   (443.5,228.23) -- (452.34,237.77) ; \draw   (452.34,228.23) -- (443.5,237.77) ;
%Straight Lines [id:da5482297103721483] 
\draw    (447.67,201.5) -- (447.89,223.25) ;
\draw [shift={(447.92,226.25)}, rotate = 269.42] [fill={rgb, 255:red, 0; green, 0; blue, 0 }  ][line width=0.08]  [draw opacity=0] (8.93,-4.29) -- (0,0) -- (8.93,4.29) -- cycle    ;
%Straight Lines [id:da2646849068818873] 
\draw    (278.83,214.33) -- (292.66,214.45) -- (298,214.16) ;
\draw [shift={(301,214)}, rotate = 176.91] [fill={rgb, 255:red, 0; green, 0; blue, 0 }  ][line width=0.08]  [draw opacity=0] (8.93,-4.29) -- (0,0) -- (8.93,4.29) -- cycle    ;
%Straight Lines [id:da8858272884608454] 
\draw    (279.5,254.33) -- (293.33,254.45) -- (300.34,254.13) ;
\draw [shift={(303.33,254)}, rotate = 177.43] [fill={rgb, 255:red, 0; green, 0; blue, 0 }  ][line width=0.08]  [draw opacity=0] (8.93,-4.29) -- (0,0) -- (8.93,4.29) -- cycle    ;
%Straight Lines [id:da603526858393002] 
\draw    (454.17,233) -- (468,233.12) -- (472.61,232.71) ;
\draw [shift={(475.6,232.44)}, rotate = 174.95] [fill={rgb, 255:red, 0; green, 0; blue, 0 }  ][line width=0.08]  [draw opacity=0] (8.93,-4.29) -- (0,0) -- (8.93,4.29) -- cycle    ;
%Straight Lines [id:da33714484383355336] 
\draw    (557,233.11) -- (587.77,198.53) ;
\draw [shift={(589.76,196.29)}, rotate = 131.66] [fill={rgb, 255:red, 0; green, 0; blue, 0 }  ][line width=0.08]  [draw opacity=0] (8.93,-4.29) -- (0,0) -- (8.93,4.29) -- cycle    ;
%Shape: Brace [id:dp6109169318758232] 
\draw   (387,70) .. controls (387,65.33) and (384.67,63) .. (380,63) -- (361,63) .. controls (354.33,63) and (351,60.67) .. (351,56) .. controls (351,60.67) and (347.67,63) .. (341,63)(344,63) -- (322,63) .. controls (317.33,63) and (315,65.33) .. (315,70) ;
%Shape: Brace [id:dp7907944950592443] 
\draw   (383.67,196.67) .. controls (383.67,192) and (381.34,189.67) .. (376.67,189.67) -- (357.67,189.67) .. controls (351,189.67) and (347.67,187.34) .. (347.67,182.67) .. controls (347.67,187.34) and (344.34,189.67) .. (337.67,189.67)(340.67,189.67) -- (318.67,189.67) .. controls (314,189.67) and (311.67,192) .. (311.67,196.67) ;
%Shape: Rectangle [id:dp005843682930295913] 
\draw  [color={rgb, 255:red, 65; green, 117; blue, 5 }  ,draw opacity=1 ][dash pattern={on 4.5pt off 4.5pt}] (22.57,9.14) -- (461.57,9.14) -- (461.57,293.14) -- (22.57,293.14) -- cycle ;
%Shape: Rectangle [id:dp8713685374650768] 
\draw  [color={rgb, 255:red, 65; green, 117; blue, 5 }  ,draw opacity=1 ][dash pattern={on 4.5pt off 4.5pt}] (461.57,9.14) -- (888.71,9.14) -- (888.71,293.14) -- (461.57,293.14) -- cycle ;
%Shape: Rectangle [id:dp23831386555690615] 
\draw  [color={rgb, 255:red, 65; green, 117; blue, 5 }  ,draw opacity=1 ][dash pattern={on 4.5pt off 4.5pt}] (888.71,9.14) -- (1094.71,9.14) -- (1094.71,293.14) -- (888.71,293.14) -- cycle ;

% Text Node
\draw (45.31,105.19) node  [font=\large]  {$x_{1}$};
% Text Node
\draw (109.67,107) node  [font=\large]  {$\mathcal{Q}( \cdot )$};
% Text Node
\draw (176.49,105.19) node  [font=\large]  {$\tilde{x}_{1}$};
% Text Node
\draw (254.67,86) node  [font=\large]  {$\mathcal{E}_{1}( \cdot )$};
% Text Node
\draw (257.07,126.09) node  [font=\large]  {$\mathcal{C}_{1}( \cdot )$};
% Text Node
\draw (449.48,64.19) node  [font=\large]  {$p_{1}$};
% Text Node
\draw (353.2,86.19) node    {$\vec{x}_{1} ,\dotsc ,\vec{x}_{1}$};
% Text Node
\draw (354.48,126.19) node    {$\tilde{c}_{1,1} ,\dotsc ,\tilde{c}_{1,L}$};
% Text Node
\draw (586.98,117.36) node  [font=\large]  {$h_{1,\ell }$};
% Text Node
\draw (654.23,177.78) node  [font=\large]  {$\sum _{k=1}^{K} h_{k,\ell } p_{k,\ell }\vec{x}_{k}\tilde{c}_{k,\ell }$};
% Text Node
\draw (517.2,105.53) node  [font=\large]  {$p_{1,\ell }\vec{x}_{1}\tilde{c}_{1,\ell }{}$};
% Text Node
\draw (752.22,128.36) node  [font=\large]  {$z$};
% Text Node
\draw (835.69,179.69) node  [font=\large,color={rgb, 255:red, 80; green, 150; blue, 220 }  ,opacity=1 ]  {$\vec{y}_{1} ,\vec{y}_{2} ,\dotsc ,\vec{y}_{L}$};
% Text Node
\draw (929.28,181.86) node  [font=\large]  {$\mathcal{T}( \cdot )$};
% Text Node
\draw (1027.31,180.03) node  [font=\large]  {$f( x_{1} ,\dotsc ,x_{K})$};
% Text Node
\draw (45.64,232.53) node  [font=\large]  {$x_{K}$};
% Text Node
\draw (110.68,234.53) node  [font=\large]  {$\mathcal{Q}( \cdot )$};
% Text Node
\draw (176.49,232.53) node  [font=\large]  {$\tilde{x}_{K}$};
% Text Node
\draw (254,213.33) node  [font=\large]  {$\mathcal{E}_{K}( \cdot )$};
% Text Node
\draw (255,253.33) node  [font=\large]  {$\mathcal{C}_{K}( \cdot )$};
% Text Node
\draw (449.81,191.53) node  [font=\large]  {$p_{K}$};
% Text Node
\draw (350.53,213.53) node  [font=\normalsize]  {$\vec{x}_{K} ,\dotsc ,\vec{x}_{K}$};
% Text Node
\draw (353.81,252.53) node  [font=\normalsize]  {$\tilde{c}_{K,1} ,\dotsc ,\tilde{c}_{K,L}$};
% Text Node
\draw (518.52,232.86) node  [font=\large]  {$p_{K,\ell }\vec{x}_{K}\tilde{c}_{K,\ell }{}$};
% Text Node
\draw (43.22,166.86) node    {$\vdots $};
% Text Node
\draw (521.22,165.86) node    {$\vdots $};
% Text Node
\draw (368.55,166.2) node    {$\vdots $};
% Text Node
\draw (172.55,166.2) node    {$\vdots $};
% Text Node
\draw (588.98,233.36) node  [font=\large]  {$h_{K,\ell }$};
% Text Node
\draw (351,44.53) node  [font=\normalsize]  {$L$};
% Text Node
\draw (347.67,171.19) node  [font=\normalsize]  {$L$};
% Text Node
\draw (245.42,27.5) node  [font=\Large] [align=left] {Transmitter};
% Text Node
\draw (677.42,29.5) node  [font=\Large] [align=left] {Channel};
% Text Node
\draw (996.42,30.5) node  [font=\Large] [align=left] {Receiver};

\end{tikzpicture}}
\caption{The proposed scheme for digital over-the-air computation by introducing ReChCompCode. %Here, the generated input value $x_k$ at user $k$ is first quantized to $\tilde{x}_k$, and then digitally modulated to $\vec{x}_k$ and distributed to corresponding transmission time slots $\tilde{c}_{k,\ell}$. The CS receives $\vec{y}_{\ell}=\sum_{k=1}^{K} h_{k,\ell}p_{k,\ell}\vec{x}_k\cdot \tilde{c}_{k,\ell}+\vec{z}_{\ell}$ in each time slot $\ell \in \{1,\ldots,L\}$. Here, $\vec{z}_{\ell}$ denotes the Gaussian white noise at each time slot, and $h_k$ denotes the channel attenuation between node $k$ and the CS. The tabular function $\mathcal{T}(\vec{y}_1,\ldots,\vec{y}_L)$ maps the value of $\vec{y}_1,\ldots,\vec{y}_L$ into the value of the function $f$ corresponding to the input values $x_k$.
}
\vspace{-10pt}
\label{fig:digital_aircomp}
\end{figure*}
% ========================

The remainder of this paper is structured as follows: Section ~\ref{sec:Systemmodel} explains the system model for the function computation problem. Section~\ref{sec:ProblemFormualtion} describes the problem of the selection process for digital modulation formats, and Section~\ref{sec:MethodDesign} proposes an efficient algorithm for accurate computation across the \ac{MAC}. In Section~\ref{sec:Num}, we present numerical experiments comparing the proposed ReChCompCode with ChannelComp and digital \ac{AirComp}. Finally, Section~\ref{sec:conclusion} concludes the paper and sketches future works.

% ================================================
\subsection{Notation}
% ================================================
Bold lower-case letter \(\bm{x}\) represents vector quantities, while bold upper-case letter \(\bm{X}\) denotes matrices. The transpose and Hermitain of a matrix \(\bm{X}\) are denoted by \(\bm{X}^\top\) and $\bm{X}^{\mathsf{H}}$, respectively. The Kronecker product is symbolized by \(\otimes\), and the Hadamard product by \(\odot\). The notation \(\bm{X} \succeq 0\) signifies that matrix \(\bm{X}\) is positive semidefinite. Lastly, \(\|X\|_2\) and \(\|X\|_{\rm F}\) represent the Euclidean norm and Frobenius norm of a matrix. For an integer $N$, we show set $\{1,2,\ldots,N\}$ by $[N]$.
\vspace{-5pt}
% =============------------==============
% ¤¤¤¤¤¤¤¤¤¤¤¤¤¤¤¤¤¤¤¤¤¤¤¤¤¤¤¤¤¤¤¤¤¤¤¤¤¤¤
\section{System Model}\label{sec:Systemmodel}
% ¤¤¤¤¤¤¤¤¤¤¤¤¤¤¤¤¤¤¤¤¤¤¤¤¤¤¤¤¤¤¤¤¤¤¤¤¤¤¤
% =============------------==============

We consider a network comprising a \ac{CP} and $K$ nodes transmitting over a shared communication channel. The main goal is to compute a function $f(x_1, \ldots, x_K)$, where $x_k \in \mathbb{R}$ represents the input value generated by node $k$. All nodes use digital modulation for transmitting their values over the MAC. In particular, node $k$ initially quantizes its value $x_k$ into $\tilde{x}_k=\mathcal{Q}(x_k) \in \mathbb{F}_Q$ using quantizer $\mathcal{Q}(\cdot)$, in which $\mathbb{F}$ represents a finite field with  $Q$ different values. Afterward, the resultant value $\tilde{x}_k$ is mapped into a digital modulation signal (complex domain) using a modulation encoder $\mathcal{E}_k(\cdot): \mathbb{F}_Q \mapsto \mathbb{C}$, resulting in $\vec{x}_k = \mathcal{E}_k(\tilde{x}_k) \in \mathbb{C}$. In this context, the values \(x_1, \ldots, x_K\) are selectively retransmitted \(L\) times, each within a designated time slot. Thus, in parallel, the quantized value \(\tilde{x}_k\) goes through channel encoder via \(\mathcal{C}: \mathbb{F}_Q \mapsto \mathbb{F}_2^{L}\), generating a length \(L\) binary sequence. This sequence serves as an indicator that \(\tilde{x}_k\) is allocated to one or more time slots, i.e., \(\tilde{\mathbf{c}}_{k} = \mathcal{C}_k(\tilde{x}_k) \in \{0,1\}^{L}\).  Subsequently, the modulated signals $\vec{x}_k$'s are simultaneously transmitted within the assigned time slot over the \ac{MAC}\footnote{We assumed that all the nodes and the \ac{CP} are perfectly synchronized. Note that the existing techniques of analog \ac{AirComp} for addressing poor synchronization, e.g., \cite{razavikia2022blind}, \cite{hellstrom2023optimal}, can be applied to digital \ac{AirComp} model as well.}. The \ac{CP} receives the superposition  of the transmitted signals at each time slot, i.e.,
% ----------------
\begin{align}
\label{eq:receivedsignal}
 \vec{y}_{\ell} = \sum\nolimits_{k=1}^{K} h_{k,\ell}p_{k,\ell}\vec{x}_k\cdot \tilde{c}_{k,\ell}+\vec{z}_{\ell},\quad \forall~\ell \in [L],   
\end{align}
% ----------------
where $\tilde{c}_{k,\ell}$ denotes element $\ell$ of vector $\tilde{{\mathbf{c}}}_{k}$, and
$h_{k,\ell} \in \mathbb{C}$ is the channel attenuation between node $k$ and \ac{CP} at time slot $\ell$. It is assumed that the wireless channel remains unchanged within every time slot. Also, $p_{k,\ell}$ is the transmit power of node $k$ at time slot $\ell$, and $\vec{z}_{\ell}$ denotes the \ac{AWGN} with zero-mean and variance $\sigma_{z}^2$ for time slot $\ell$.

According to the optimal power control policy suggested by AirComp literature~\cite{cao2020optimal}, we consider the selection of transmit power as
the inverse of the channel, i.e., $p_{k,\ell}=h_{k,\ell}^*/|h_{k,\ell}|^2$. Note that insufficient power nodes would only result in weaker signal strength.  With this, we can express \eqref{eq:receivedsignal} as follows:
% ----------------
\begin{equation}
\label{eq:nofading}
\vec{y}_{\ell} = \sum\nolimits_{k=1}^{K} \vec{s}_{k,\ell} + \vec{z}_{\ell}, \quad \forall~\ell \in [L],
\end{equation}
% ----------------
 where $\vec{s}_{k,\ell}:=  \vec{x}_k \cdot \tilde{c}_{k,\ell}$ denotes the transmitted signal at time slot $\ell$ by node $k$ for $\ell \in [L]$ and $k\in [K]$.  Because every transmitted  $\vec{s}_{k,\ell}$ possesses finite possible constellation points, the received signal $\vec{y}_{\ell}$ acquires a finite constellation diagram. To compute the desired function $f$, we employ a Tabular mapping $\mathcal{T}$ on the sequence of $\vec{y}_{1}, \ldots, \vec{y}_{L}$ to obtain the corresponding output.  The entire procedure of this channel computing coding is illustrated in Fig.~\ref{fig:digital_aircomp}.

In the next section, we design encoders and decoders to enable performing the computation of the desired function $f$ over the \ac{MAC}.

\vspace{-5pt}
% =============------------==============
% ¤¤¤¤¤¤¤¤¤¤¤¤¤¤¤¤¤¤¤¤¤¤¤¤¤¤¤¤¤¤¤¤¤¤¤¤¤¤¤
\section{Problem Formulation}\label{sec:ProblemFormualtion}
% ¤¤¤¤¤¤¤¤¤¤¤¤¤¤¤¤¤¤¤¤¤¤¤¤¤¤¤¤¤¤¤¤¤¤¤¤¤¤¤
% =============------------==============

This section introduces the proposed channel computing code termed ReChCompCode, which enables robust computation over the \ac{MAC}. 

Towards acquiring a robust and efficient computation, we need to jointly design modulation encoder $\mathcal{E}_k$ and channel encoder $\mathcal{C}_k$ for all users $k \in [K]$ to avoid any destructive overlaps of constellation points in the receiver side. Therefore, the tabular function $\mathcal{T}$ can uniquely map the constellation points of  $\Vec{y}_{1}, \ldots, \Vec{y}_{L}$ to the corresponding outputs of the function $f$. Specifically, consider a noiseless \ac{MAC}. Then, let $f^{(i)}$ and $f^{(j)}$ be two output values from the range of function $f(x_1,\ldots,x_K)$ for two sets of input values $x_1^{(i)},\ldots,x_K^{(i)}$ and  $x_1^{(j)},\ldots,x_K^{(j)}$, respectively. 
Then, the induced constellation points at time slot $\ell$ can be expressed by $\Vec{v}_{\ell}^{(i)}:= \sum_{k}\vec{s}_{k,\ell}^{(i)}$
and $\Vec{v}_{\ell}^{(j)}:= \sum_{k}\vec{s}_{k,\ell}^{(j)}$. To have a valid computation, the sequence of the induced points $\bm{v}^{(i)}:=[\Vec{v}_{1}^{(i)},\ldots,\Vec{v}_{L}^{(i)}]$ and $\bm{v}^{(j)}:=[\Vec{v}_{1}^{(j)},\ldots,\Vec{v}_{L}^{(j)}]$ need to satisfy the following constraint~\cite{razavikia2023computing}:
\begin{align}
    \label{eq:CompCond}
    {\rm if }~f^{(i)}\neq f^{(j)}~\Rightarrow~\bm{v}^{(i)} \neq \bm{v}^{(j)},\quad \forall (i,j)\in [M]^2,
\end{align}
where $[M]^2 = [M]\times [M]$ is the Cartesian product and  $M$ is the cardinality of the range of function $f$. The constraint in~\eqref{eq:CompCond} ensures that the tabular function $\mathcal{T}$ can correctly map $\Vec{v}^{(i)}$ to output value $f^{(i)}$ based on the design of $\mathcal{E}_k$ and $\mathcal{C}_k$. Therefore,  the problem formulation involves designing the constellation points in a way that ensures condition~\eqref{eq:CompCond} is satisfied.

% ================== Example =============================
\subsection{Illustrative Example}

For better insight, we illustrate the mechanism by a simple case. Consider a network involving $K=4$ nodes, where they employ a simple digital modulation, such as quadrature phase shift keying modulation, to compute the product function, i.e.,  $f(x_1,x_2,x_3,x_4) = x_1x_2x_3x_4$ over $L=2$ time slots. The input value $x_k \in \{1, 2, 3, 4\}$ is quantized into a two-bit representation, denoted as $\tilde{x}_k \in \{00, 01, 10, 11\}$. Each quantized value can be encoded as $\vec{x}_k \in \{1,-1,i,-i\}$. In the case of a noiseless \ac{MAC}, when the four nodes transmit the same signals in both time slots, a conflict occurs at point $0$ in the constellation, where point $0$ corresponds to three different output values, i.e., \(\{4,24,144\}\).  However, this overlapping can be avoided by performing the computation over two separate time slots using the channel encoder defined as follows. 
% ----------------
\begin{equation}
\tilde{c}_{k,0} = \left\{
\begin{aligned}
1, & \quad \text{if } \tilde{x}_k \in \{00,10\} \\
0, & \quad \text{if } \tilde{x}_k \in \{01,11\}
\end{aligned}
\right.  \\
,\tilde{c}_{k,1} = \left\{
\begin{aligned}
0, & \quad \text{if } \tilde{x}_k \in \{00,10\} \\
1, & \quad \text{if } \tilde{x}_k \in \{01,11\}
\end{aligned}
\right.
\end{equation}
% ----------------

Table \ref{illustrative example} provides an explicit comparison. As the received signals \(\vec{y}_{\ell} = \sum_{k=1}^{K} \vec{x}_k\cdot \tilde{c}_{k,\ell}\) for \(\ell \in \{1,2\}\) are distinct over time slots, the corresponding function outputs can be distinguished by using a tabular function $\mathcal{T}$.

In the following subsection, we propose an optimization problem to jointly design the modulation encoder and channel computing encoder for a general function computation. 
% % ========================
% \begin{table}[t]
% \caption{Difference between with and without channel computing coding.}
% \begin{center}
% \begin{tabular}{|c|c|c|c|}
% \hline
% $x_1$ $x_2$ & $x_1x_2x_3x_4$ & $\vec{y}_{\ell}=\sum_{k=1}^{K} \Vec{x}_k$ & $\vec{y}_{\ell } = \sum_{k=1}^{K} \Vec{x}_k \tilde{c}_{k,\ell}$ \\
% \hline
% 1 \quad 1 & 1 & 2 \quad 2 & 2 \quad 0 \\
% 1 \quad 2  & \textcolor{red}2 & \textcolor{red}0 \quad \textcolor{red}0 & \textcolor{bluegray}{1 \quad -1}\\
% 1 \quad 3  & 3 & 1+i \quad 1+i & 1+i \quad 0\\
% 1 \quad 4  & 4 & 1-i \quad 1-i & 1 \quad -i \\
% 2 \quad 2  & 4 & -2 \quad -2 & 0 \quad -2 \\
% 2 \quad 3  & 6 & -1+i \quad -1+i & i \quad -1 \\
% 2 \quad 4  & 8 & -1-i \quad -1-i & -1-i \quad 0 \\
% 3 \quad 3  & 9 & 2i \quad 2i & 2i \quad 0 \\
% 3 \quad 4  & \textcolor{red}{12} & \textcolor{red}0 \quad \textcolor{red}0 & \textcolor{bluegray}{i \quad -i} \\
% 4 \quad 4  & 16 & -2i \quad 2i & 0 \quad -2i \\
% \hline
% \multicolumn{4}{p{0.9\linewidth}}{An intuitive example to show that utilizing channel computing coding can avoid overlap in the resultant constellation diagram.}
% %\multicolumn{4}{l}{$^{\mathrm{a}}$Sample of a Table footnote.}
% \label{illustrative example}
% \end{tabular}

% \label{tab1}
% \end{center}
% \end{table}

\begin{table}[t]
\caption{Difference between with and without ReChCompCode.}
\begin{center}
\begin{tabular}{|c|c|c|c|}
\hline
$x_1$ $x_2$ $x_3$ $x_4$ & $x_1x_2x_3x_4$ & $\vec{y}_{\ell}=\sum_{k=1}^{K} \Vec{x}_k$ & $\vec{y}_{\ell } = \sum_{k=1}^{K} \Vec{x}_k \tilde{c}_{k,\ell}$ \\
\hline
1 \quad 1 \quad 1 \quad 1 & 1 & 4 \quad 4 & 4 \quad 0 \\
1 \quad 1 \quad 2 \quad 2 & \textcolor{red}4 & \textcolor{red}0 \quad \textcolor{red}0 & \textcolor{bluegray}{2 \quad -2} \\
1 \quad 2 \quad 2 \quad 2  & 8 & -2 \quad -2 & 1 \quad -3 \\
1 \quad 2 \quad 2 \quad 3  & 12 & -1+i\quad -1+i & 1+i \quad -2 \\
1 \quad 2 \quad 2 \quad 4  & 16 & -1-i\quad -1-i & 1 \quad -2-i \\
1 \quad 2 \quad 3 \quad 4 & \textcolor{red}{24} & \textcolor{red}0 \quad \textcolor{red}0 & \textcolor{bluegray}{1+i \quad -1-i}\\
2 \quad 2 \quad 2 \quad 2  & 16 & -4\quad -4 & 0 \quad -4 \\
%2 \quad 3 \quad 3 \quad 3  & 54 & -1+3i\quad -1+3i & 3i \quad -1 \\
2 \quad 3 \quad 3 \quad 4  & 72 & -1+i\quad -1+i & 2i \quad -1-i \\
3 \quad 3 \quad 3 \quad 3  & 81 & 4i\quad 4i & 4i \quad 0 \\
3 \quad 3 \quad 4 \quad 4 & \textcolor{red}{144} & \textcolor{red}0 \quad \textcolor{red}0 & \textcolor{bluegray}{2i \quad -2i}\\
4 \quad 4 \quad 4 \quad 4  & 256 & -4i\quad -4i & 0 \quad -4i \\
\hline
\multicolumn{4}{p{0.9\linewidth}}{An intuitive example to show that repeating transmissions with ChannelComp can avoid overlapping of points in the constellation diagram at the receiver, thus making it possible correct computations.}
%\multicolumn{4}{l}{$^{\mathrm{a}}$Sample of a Table footnote.}
\label{illustrative example}
\end{tabular}

\label{tab1}
\end{center}
\vspace{-10pt}
\end{table}
% % ========================

% ==============================================
\subsection{ReChCompCode Design}
% ================================================
In the following discussion, we will use matrix notation to represent points in the function domain for improved clarity. For each node $k$, we define  $\bm{x}_k := [x_k^{(1)},\ldots,x_k^{(Q)}]^{\mathsf{T}} \in \mathbb{C}^{Q \times 1}$ as modulation vector containing all $Q$ possible constellation points created by the modulation encoder $\mathcal{E}_k(\cdot)$. Similarly, the channel matrix of node $k$ can be defined as $\bm{C}_k \in \{0,1\}^{L\times Q}$ whose $[\bm{C}_k]_{(\ell,q)} := c_{k,\ell}^{(q)} \in \{0,1\}$ denotes  whether the modulated signal $x_k^{(q)}$ is transmitted in time slot $\ell$. We further define  $\bm{x} := [\bm{x}_1^{\mathsf{T}}, \ldots, \bm{x}_K^{\mathsf{T}}]^{\mathsf{T}} \in \mathbb{C}^{N \times 1}$ and the block matrix $\bm{C} := [\bm{C}_1, \ldots, \bm{C}_K]^{\mathsf{T}} \in \{0,1\}^{N \times L}$, where $N:= Q K$.  Then, all the possible transmitted signals can be expressed by $\bm{S}: = \bm{X} \odot \bm{C}$, in which $\bm{X} = \bm{x} \otimes  \mathds{1}_{L}^{\mathsf{T}} \in \mathbb{C}^{N\times L}$ and $\mathds{1}_{L}$ is a $L\times 1$ column vector of 1s. As a result, we can express all possible constellation points over the \ac{MAC} for all $L$ time slots as 
%----------------
\begin{equation}
\bm{V} = \bm{A} \bm{S}\in  \mathbb{C}^{M \times L}, 
\end{equation}
%----------------
where $\bm{A} \in  \{0, 1\}^{M \times N}$ is a binary matrix whose $i$-th row is a binary vector with the same support as input values corresponding to $f^{(i)}$. Indeed, $\bm{A}$ selects elements of $\bm{S}$ such that column $\ell$ of matrix $\bm{V}$ consists of all possible constellation points experienced by the \ac{CP} at time slot $\ell$. 

To hold an error-free computation over the \ac{MAC}, the necessary and sufficient condition is to have distinct sequences of constellation points for corresponding different function output values. In particular, for a pair $f^{(i)}$ and $f^{(j)}$, the resulting vectors $\bm{v}^{(i)}$ and $\bm{v}^{(j)}$ must satisfy the following constraint:
%------------
\begin{align}
    \label{eq:CondMain}
    \| \bm{v}^{(i)} - \bm{v}^{(j)}\|_{2}^2 \geq  \epsilon |f^{(i)} - f^{(j)}|^2, \quad \forall (i,j)\in [M]^2,
\end{align}
%------------
where $\bm{v}^{(i)} \in \mathbb{C}^{L}$ involves all $L$ induced constellation points corresponding to $f^{(i)}$, and $\epsilon$ is a positive constant that needs to be adjusted based on the input function values and the variance of the noise at the receiver as given in~\cite{saeed2023ChannelComp}. The inequality in~\eqref{eq:CondMain} ensures that resultant constellation points are distinguishable for tabular function $\mathcal{T}$. Certainly, the modulation vector that can satisfy this constraint is not unique. To attain a robust and power-efficient channel computing coding, we pose the following optimization problem:
%------------
\begin{align}
\nonumber
 \mathcal{P}_{0} :=\min_{\bm{x},\bm{C}} \quad & ||\bm{S}||_{\rm F} \\ \label{eq:smooth original problem}
{\rm s.t.} \quad&  \| (\bm{a}_{i} - \bm{a}_{j})^{\mathsf{T}}\bm{S}\|_{\rm F}^2 \geq  \epsilon |f^{(i)} - f^{(j)}|^2,  \\ \nonumber
& (i,j) \in [M]^2,  \quad \|\bm{x}\|_2^2 \leq P_{\max},
\end{align}
%------------
where $\bm{a}_i^{\mathsf{T}}$ denotes the $i$-th row of matrix $\bm{A}$ and $\bm{a}_i^{\mathsf{T}}\bm{S} = \bm{v}^{(i)}$. Problem $\mathcal{P}_0$ jointly optimize the modulation vector $\bm{X}$ and channel vector $\bm{C}$ while guaranteeing that computation constraint holds. By solving Problem $\mathcal{P}_0$, we acquire the optimum $\bm{X}^*$ and $\bm{C}^*$, thereafter, we obtain the modulation encoders $\mathcal{E}_1(\cdot), \ldots, \mathcal{E}_K(\cdot)$  and channel encoders $\mathcal{C}_1(\cdot), \ldots, \mathcal{C}_K(\cdot)$, respectively, that map the input values to the resultant modulation vector and channel vector.

From~\cite{saeed2023ChannelComp}, we note that  Problem $\mathcal{P}_0$ is NP-hard due to the nonconvex quadratic constraints in \eqref{eq:smooth original problem}. Hence, we provide an efficient algorithm to compute an approximate solution in the next section.   

% ================================================
\section{Modulation and Channel Coding Co-Design }\label{sec:MethodDesign}
% ================================================

In this section, to find an approximate solution for Problem $\mathcal{P}_0$, we first pose an alternating optimization problem, and then we solve it and analyze its optimality. Based on such a solution, we establish a receiver architecture for recovering the function output from the received constellation vector. 
% ================================================
\subsection{Approximate Solution via alternating Minimization }
% ================================================
The proposed optimization by Problem $\mathcal{P}_0$ is an ill-posed problem due to the Hadamard product of $\bm{X}$ and $\bm{C}$. To circumvent this, we approximate the objective function with the following upper bound:
%----------------
\begin{align}
    \nonumber
    \|\bm{S} \|_{\rm F}  & =  \|\bm{X} \odot \bm{C} \|_{\rm F} \leq \|\bm{X}\|_{\rm F}\|\bm{C}\|_{\rm F},\\
      & \leq \frac{1}{2}(\|\bm{X}\|_{\rm F}^2 + \|\bm{C}\|_{\rm F}^2) = \frac{1}{2}( L \|\bm{x}\|_2^2 + \sum\nolimits_{\ell=1}^L\|\bm{c}_{\ell}\|_2^2),\label{eq:INequality}
\end{align}
%----------------
where $\bm{c}_{\ell}$ denotes column $\ell$ of matrix $\bm{C}$ for $\ell  \in [L]$.  Therefore, using \eqref{eq:INequality}, we minimize the upper bound of the objective function in \eqref{eq:smooth original problem} as follows
%----------------
\begin{subequations} \label{eq:relaxed original problem} 
\begin{align} 
\nonumber
  \mathcal{P}_{1} :=\min_{\bm{x},\bm{C}} \quad & L||\bm{x}||_2^2 + \sum\nolimits_{\ell=1}^L ||\bm{c}_{\ell}||^2_2  \\ 
\text{s.t.} \quad & \sum\nolimits_{\ell=1}^L |(\bm{a}_i-\bm{a}_j)^\top(\bm{x}\odot\bm{c}_{\ell})|^2 \geq \Delta f_{i,j}, \\
& \quad \|\bm{x}\|_2^2 \leq P_{\max}, \label{eq:relaxed power constraint}
\end{align}
\end{subequations}
%----------------
where $\Delta f_{i,j}:=\epsilon|f^{(i)}-f^{(j)}|^2$ and the digital vector as $\bm{x} \in \mathbb{C}^{N \times 1}$ and the time slot vector as $\bm{c}_{\ell} \in \{0,1\}^{N \times 1}$.

To explore the transmission efficiency, we can find the minimum value of total time slot $L$ from Problem $\mathcal{P}_{1}$.
\begin{prop}\label{pro:Lmin}
For a given function $f$, to obtain a feasible solution for Problem $\mathcal{P}_1$   satisfying the power constraint~\eqref{eq:relaxed power constraint}, the number of time slots $L$ must be greater than or equal to the following lower bound. 
%----------------
\begin{equation}
\begin{split}
L_{\min} \geq \frac{\max_{i,j} \Delta f_{i,j} \cdot N}{P_{\max}}.
\end{split}
\end{equation}
%----------------
\end{prop}

Given the lower bound $L_{\min}$, we set a fixed value for $L$ to solve Problem $ \mathcal{P}_{1}$. By using an alternating approach, Problem $ \mathcal{P}_{1}$ can be split into two subproblems by iteratively optimizing decision variables $\bm{x}$ and $\bm{C}$. More precisely, given $\bm{c}_{\ell}^{(n-1)}$ at iteration $n$,  we optimize over $\bm{x}$ in the following subproblem.
%----------------
\begin{subequations}
\begin{align}
    \nonumber
    \bm{x}^{(n)} & = \arg \min_{\bm{x}} \ L\|\bm{x}\|^2_2 \\
    \text{s.t.} &\ \sum\nolimits_{\ell=1}^L |(\bm{a}_i-\bm{a}_j)^\top(\bm{x}\odot\bm{c}_{\ell}^{(n-1)})|^2 \geq \Delta f_{i,j}, \\
    & \quad \|\bm{x}\|_2^2 \leq P_{\max}.
\label{eq:CVX}
\end{align}
\end{subequations}
%----------------

For the variable $\bm{c}_{\ell}$,  we have
%----------------
\begin{equation}
\begin{aligned}
    \bm{c}_{\ell}^{(n)} & = \arg \min_{\bm{c}_{\ell}} \ \sum\nolimits_{\ell=1}^L||\bm{c}_{\ell}||^2_2 \\
    \text{s.t.} &\ \sum\nolimits_{\ell=1}^L |(\bm{a}_i-\bm{a}_j)^\top(\bm{x}^{(n)}\odot\bm{c}_{\ell})|^2 \geq \Delta f_{i,j}.
\label{eq:MBO}
\end{aligned}
\end{equation}
%----------------

For the first subproblem in \eqref{eq:CVX}, invoking the self-adjoint property of the Hadamard product, i.e., $|\bm{x}^\top(\bm{y}\odot\bm{z})|^2 = |\left(\bm{x}\odot\bm{z}\right)^\top\bm{y}|^2$, we reformulate the constraint as:
%----------------
\begin{equation}
\begin{split}
\min_{\bm{x}} &\quad  L\| \bm{x}\|^2_2   \\
\text{s.t.} &\quad  \sum\nolimits_{\ell=1}^L \left|\left((\bm{a}_i-\bm{a}_j)\odot\bm{c}_{\ell}^{(n-1)}\right)^\top\bm{x} \right|^2 \geq \Delta f_{i,j}, \\
& \quad \|\bm{x}\|_2^2 \leq P_{\max}.
\end{split}
\label{eq:reformulated_CVX}
\end{equation}
%----------------   

The optimization problem in \eqref{eq:reformulated_CVX} is a \ac{QCQP}, which is non-convex and NP-hard~\cite{sidiropoulos2006transmit}. One possible way to handle this problem is to use the lifting trick \cite{vandenberghe1996semidefinite}, where the non-convexity appears as a rank constraint.  
In particular, we rewrite the problem in terms of a new lifted matrix variable $\bm{W}:= \bm{x} \bm{x}^{\mathsf{H}}$ as \eqref{eq:LiftSDP}.  
%-------------
\begin{subequations}
\label{eq:LiftSDP}
\begin{align}
\nonumber
\bm{W}^{(n)} = \arg \min_{\bm{W}}  \quad & {\rm Tr}(\bm{W})   \\ \label{eq:lifting CVX}
{\rm s.t.} \quad & {\rm Tr}(\bm{W} \cdot \bm{B}_{i,j}^{(n-1)}) \geq \Delta f_{i,j},\\
 & \bm{W} \succeq \bm{0}, \quad {\rm rank}(\bm{W})=1,  \label{eq:lift-rank} \\
& \text{Tr}(\mathbf{W}) \leq P_{\max},
\end{align}
\end{subequations}
%-------------
where $$\bm{B}_{i,j}^{(n-1)}=\sum\nolimits_{\ell=1}^L((\bm{a}_i-\bm{a}_j)\odot\bm{c}_{\ell}^{(n-1)})((\bm{a}_i-\bm{a}_j)\odot\bm{c}_{\ell}^{(n-1)})^\top.$$ Next, dropping the rank-one constraint in \eqref{eq:lift-rank}, the optimization problem in \eqref{eq:LiftSDP} becomes \ac{SDP} and can be solved by convex solver tolls such as CVX~\cite{grant2014cvx}. After solving the \ac{SDP} problem, $\bm{x}^{(n)}$ is obtained via Cholesky decomposition of $\bm{W}^{(n)}$ with rank one property. In case  $\bm{W}^{(n)}$ possess a rank higher than one, an approximate solution $\bm{x}^{(n)}$ can be recovered using the Gaussian randomization method~\cite{luo2010semidefinite}.

Similarly, for the subproblem in \eqref{eq:MBO}, since $\bm{c}_{\ell}$ is binary, the quadratic objective $\|\bm{c}_{\ell}\|_2^2$ is equivalent to a linear one as $\sum_{l=1}^L\bm{w}^\top\bm{c}_{\ell}$, where $\bm{w}$ is $\mathds{1}_{L}$. Moreover, using self-adjoint property of the Hadamard product, we can rewrite~\eqref{eq:MBO} as 
%----------------
\begin{subequations}
\label{eq:MBO_reformulated2}
\begin{align}
\nonumber
 \bm{c}_{\ell}^{(n)} = \arg \min_{\bm{c}_{\ell}} \quad & \sum\nolimits_{\ell=1}^L\bm{w}^\top\bm{c}_{\ell} \\
{\rm s.t.} \quad & \sum\nolimits_{\ell=1}^L\left|\bm{d}_{i,j}^{(n)\mathsf{H}} \bm{c}_{\ell}\right|^2 \geq \Delta f_{i,j},
\end{align}
\end{subequations}
where $\bm{d}_{i,j}^{(n)}: = (\bm{a}_i - \bm{a}_j) \odot \bm{x}^{(n)}$.  The optimization problem in \eqref{eq:MBO_reformulated2} is a \ac{MIP} problem with linear objective and quadratic constraints. This category of problems can be solved through solvers like Gurobi \cite{pedroso2011optimization} by using the branch and bound method~\cite{lawler1966branch}.
The entire procedure of the alternating minimization approach is summarized in Algorithm~\ref{Alg:HIT}.

% ========================
\begin{algorithm}[!t]\label{Alg:HIT}
    \caption{ReChCompCode Algorithm}
    \KwData{Initial values: $\bm{x}^{(0)}$, $\bm{C}^{(0)}$}
    \KwResult{Optimized values of $\bm{x}$ and $\bm{C}$}
    
     Set  Number of iterations $N$ and initialize iteration counter $n = 1$\;

    \While{not converged and $n \leq N$}{
        Update $\bm{x}$:
        
        \Indp
        Obtain $\bm{W}^{(n)}$ by solving \eqref{eq:LiftSDP}. \\
        
        Compute $\bm{x}^{(n)}$ via Cholesky decomposition of $\bm{W}^{(n)}$.
        
        \Indm
        
        Update $\bm{C}$:
        
        \Indp
        
        Solve the subproblem (\ref{eq:MBO}) to obtain $\bm{C}^{(n)}$ given   $\bm{x}^{(n)}$;

        \Indm
        
        Increment the iteration counter: $n \leftarrow n + 1$\;
        
        Check for convergence:
        
        \Indp
        If $n=N$ or $L||\bm{x}^{(n)}-\bm{x}^{(n-1)}||^2 + \sum_{\ell=1}^L ||\bm{c}_{\ell}^{{(n)}}-\bm{c}_{\ell}^{{(n-1)}}||^2 \leq \delta$, stop iterating\;
        \Indm
    }
    
    \textbf{Output:} the optimized values of $\bm{x}^{*}$ and $\bm{C}^{*}$.
 
\end{algorithm}

% ========================

%====================================
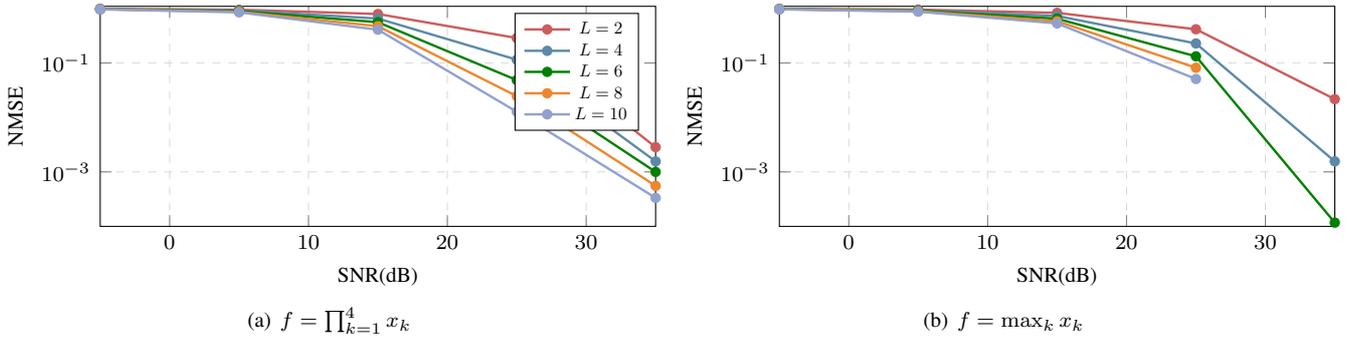
\begin{figure*}[!t]
\centering
\subfigure[$f = \prod_{k=1}^4x_k$]{   \label{fig:four_node_product_SNR}
\begin{tikzpicture}
    \begin{axis}[
        xlabel = {SNR(dB)},
        ylabel = {NMSE},
        label style={font=\footnotesize},
        width=0.49\textwidth,
        height=4.5cm,
        xmin=-5, xmax=35,
        ymin=1e-4, ymax=1.1,
        legend style={nodes={scale=0.65, transform shape}, at={(0.008,0.0006)} },
        ticklabel style = {font=\footnotesize},
        legend pos=north east,
        ymajorgrids=true,
        xmajorgrids=true,
        grid style=dashed,
        grid=both,
        ymode = log,
        grid style={line width=.1pt, draw=gray!10},
        major grid style={line width=.2pt,draw=gray!30},
    ]
    \addplot[%smooth,
             thin,
        color=chestnut,
        mark=*,
        line width=0.9pt,
        mark size=1.5pt,
        ]
    table[x=SNR,y=L2]
    {Data/Sim.dat};
    \addplot[ %smooth,
              %thin,
            color=airforceblue,
            mark=*,
            line width=0.9pt,
            mark size=1.5pt,
            ]
    table[x=SNR,y=L4]
    {Data/Sim.dat};
    \addplot[ %smooth,
             %thin,
        color=cssgreen,
        mark=*,
        line width=0.9pt,
        mark size=1.5pt,
        ]
    table[x=SNR,y=L6]
    {Data/Sim.dat};
    \addplot[ %smooth,
             %thin,
        color=cadmiumorange,
        mark=*,
        line width=0.9pt,
        mark size=1.5pt,
        ]
    table[x=SNR,y=L8]
    {Data/Sim.dat};
    \addplot[ %smooth,
             %thin,
        color=ceil,
        mark=*,
        line width=0.9pt,
        mark size=1.5pt,
        ]
    table[x=SNR,y=L10]
    {Data/Sim.dat};
    \legend{$L=2$, $L=4$,$L=6$, $L=8$,$L=10$};
\end{axis}
\end{tikzpicture}
}\subfigure[$f=\max_{k}x_k$]{\label{fig:four_node_max_SNR}
\begin{tikzpicture}
    \begin{axis}[
        xlabel = {SNR(dB)},
        ylabel = {NMSE},
        label style={font=\footnotesize},
        width=0.49\textwidth,
        height=4.5cm,
        xmin=-5, xmax=35,
        ymin=1e-4, ymax=1.1,
        legend style={nodes={scale=0.65, transform shape}, at={(0.3,0.85)}},
        ticklabel style = {font=\footnotesize},
        legend pos=north east,
        ymajorgrids=true,
        xmajorgrids=true,
        grid style=dashed,
        grid=both,
        ymode = log,
        grid style={line width=.1pt, draw=gray!10},
        major grid style={line width=.2pt,draw=gray!30},
    ]
    \addplot[%smooth,
             thin,
        color=chestnut,
        mark=*,
        line width=0.9pt,
        mark size=1.5pt,
        ]
    table[x=SNR,y=L2_m_4]
    {Data/Sim.dat};
    \addplot[ %smooth,
              %thin,
            color=airforceblue,
            mark=*,
            line width=0.9pt,
            mark size=1.5pt,
            ]
    table[x=SNR,y=L4_m_4]
    {Data/Sim.dat};
    \addplot[ %smooth,
             %thin,
        color=cssgreen,
        mark=*,
        line width=0.9pt,
        mark size=1.5pt,
        ]
    table[x=SNR,y=L6_m_4]
    {Data/Sim.dat};
    \addplot[ %smooth,
             %thin,
        color=cadmiumorange,
        mark=*,
        line width=0.9pt,
        mark size=1.5pt,
        ]
    table[x=SNR,y=L8_m_4]
    {Data/Sim.dat};
    \addplot[ %smooth,
             %thin,
        color=ceil,
        mark=*,
        line width=0.9pt,
        mark size=1.5pt,
        ]
    table[x=SNR,y=L10_m_4]
    {Data/Sim.dat};
    % \legend{$L=2$, $L=4$,$L=6$, $L=8$,$L=10$};
\end{axis}
\end{tikzpicture}}
  \caption{Performance of ReChCompCode under different SNRs in terms of NMSE averaged over $N_s=100$. The input values are $x_k=\{0,1,2,3\}$ and the desired functions are $f = \prod_{k=1}^4x_k$ and $f=\max_{k}x_k$ with $K=4$ nodes.}
\end{figure*}

%====================================

\subsection{Optimality Analysis}

We analyze the optimality of Algorithm~\ref{Alg:HIT}, where we use an alternating minimization approach for solving Problem $\mathcal{P}_1$. 
First, considering the \ac{QCQP} problem in~\eqref{eq:LiftSDP}, 
since the rank constraint is removed, the solution $\bm{W}^*$ is suboptimal. However, if $\bm{W}^*$ becomes a rank-one matrix, it is feasible to~\eqref{eq:LiftSDP}, and in fact, optimal solution. 
In addition, the \ac{MIP} problem in \eqref{eq:MBO_reformulated2} has non-convex quadratic constraints. By translating these quadratic constraints into bilinear ones, we can apply the reformulation linearization technique and an implicit enumeration strategy, and the global optimum can be obtained, as outlined in~\cite{sherali2013reformulation}. The convergence of the alternating minimization algorithm for the bilinear constraint is studied in \cite{liu2023fast}. Overall, by relaxing the objective function in Problem $\mathcal{P}_1$, the final solution $\bm{x}^{*}$ and $\bm{C}^{*}$ serves as an approximate solution to Problem $\mathcal{P}_0$.

% =====================================
\subsection{Designing Tabular Function}
% =====================================

The tabular function $\mathcal{T}$ is uniquely determined by the decision boundaries based on the reshaped constellation points over the \ac{MAC}. Indeed,  the \ac{CP} uses a set of constellation points \(\bm{v}^{(i)} = [\vec{v}^{(i)}_{1}, \ldots, \vec{v}^{(i)}_{L}]^{\mathsf{T}} \in \mathbb{C}^{L \times 1}\) to decode the function output $f^{(i)}$, where \(\vec{v}_{\ell}^{(i)}=\sum_{k=1}^K \vec{x}_{k} \cdot \tilde{c}_{k,\ell}\). Then, the problem becomes finding the corresponding points of $\vec{\bm{v}}^{(i)}$ when $\bm{y}$ is received. Following similar steps as in \cite{razavikia2023computing}, one can show the maximum likelihood estimator results in the problem below.
% --------------
\begin{align}
\hat{f}^{(i)}=\arg \min_{i}||\bm{y} - \bm{v}^{(i)}||_2^2.
\end{align}
% --------------
The decoder generates the constellation diagram, which consists of all possible constellation points $\{\vec{v}_{\ell}^{(1)},\ldots,\vec{v}_{\ell}^{(M)}\}$ along with their corresponding Voronoi cells $\{\mathcal{V}_{1,\ell},\ldots,\mathcal{V}_{M,\ell}\}$. Each Voronoi cell contains the points in the signal space that are closer to a specific constellation point than to any other.
The desired function $\hat{f}$ for all time slots is given by $\hat{f}=\sum_{j=1}^M\mathcal{T}_{j}(\bm{v})$, where $\mathcal{T}_{j}(\cdot)$ is an indicator function: 
% --------------
\begin{align}
\mathcal{T}_{j}(\bm{v}) := \begin{cases}
    \hat{f}^{(j)}, & \quad \text{if } \vec{v}_{\ell} \in \mathcal{V}_{j,\ell}, \ell \in [L], \\
0, & \quad \text{o.w.} 
\end{cases}    
\end{align}
% --------------
For the points that the overlapping is not resolved, i.e., corresponding constraints in~\eqref{eq:relaxed original problem} are not satisfied, their Voronoi cells merge into one Voronoi cell, and the decoder has to assign only one number as the output of two different output values for the desired functions, e.g., the average of output values corresponding to the Voronoi cells~\cite{saeed2023ChannelComp}.  

In the next section, we check the performance of the proposed ReChCompCode over the \ac{MAC}.

\section{Numerical Experiments}\label{sec:Num}

In this section, we assess the performance of ReChCompCode for computing different functions over the \ac{MAC}. Moreover, we make a comparison of the ReChCompCode with ChannelComp~\cite{razavikia2023computing} and the retransmission strategy for digital \ac{AirComp} proposed in~\cite{goldenbaum2014nomographic}. More precisely, we use a naive repetition strategy for the ChannelComp~\cite{razavikia2023computing} and \ac{AirComp}, where the \ac{CP} average the received signal over multiple transmission as the estimation value for the output of the function.

\subsection{Performance of ReChCompCode}

%====================================
\begin{figure}[!t]
\centering
\subfigure[$f = \sum_{k=1}^fx_k$]{\label{fig:four_node_sum_NMSE}
\begin{tikzpicture}
    \begin{axis}[
        xlabel = {L},
        ylabel = {NMSE},
        label style={font=\footnotesize},
        width=0.48\textwidth,
        height=6cm,
        xmin=0, xmax=100,
        ymin=5e-4, ymax=0.85,
        legend style={nodes={scale=0.65, transform shape}, at={(0.3,0.85)}},
        ticklabel style = {font=\footnotesize},
        legend pos=north east,
        ymajorgrids=true,
        xmajorgrids=true,
        grid style=dashed,
        grid=both,
        ymode = log,
        grid style={line width=.1pt, draw=gray!10},
        major grid style={line width=.2pt,draw=gray!30},
    ]
    \addplot[%smooth,
             thin,
        color=chestnut,
        mark=*,
        line width=0.9pt,
        mark size=1.5pt,
        ]
    table[x=L,y=AirComp]
    {Data/Sim_L.dat};
    \addplot[ %smooth,
              %thin,
            color=airforceblue,
            mark=*,
            line width=0.9pt,
            mark size=1.5pt,
            ]
    table[x=L,y=ChannelComp]
    {Data/Sim_L.dat};
    \addplot[ %smooth,
             %thin,
        color=cssgreen,
        mark=*,
        line width=0.9pt,
        mark size=1.5pt,
        ]
    table[x=L,y=ChannelCoding]
    {Data/Sim_L.dat};
    \legend{Digital AirComp~\cite{goldenbaum2014nomographic}, ChannelComp~\cite{saeed2023ChannelComp}, ReChCompCode};
\end{axis}
\end{tikzpicture}
  }
  \subfigure[$f=\prod_{k=1}^4x_k$]{\label{fig:four_node_prod_NMSE}
\centering
\begin{tikzpicture}
    \begin{axis}[
        xlabel = {L},
        ylabel = {NMSE},
        label style={font=\footnotesize},
        width=0.48\textwidth,
        height=6cm,
        xmin=0, xmax=100,
        ymin=5e-5, ymax=0.4,
        legend style={nodes={scale=0.65, transform shape}, at={(0.3,0.85)}},
        ticklabel style = {font=\footnotesize},
        legend pos=north east,
        ymajorgrids=true,
        xmajorgrids=true,
        grid style=dashed,
        grid=both,
        ymode = log,
        grid style={line width=.1pt, draw=gray!10},
        major grid style={line width=.2pt,draw=gray!30},
    ]
    \addplot[%smooth,
             thin,
        color=chestnut,
        mark=*,
        line width=0.9pt,
        mark size=1.5pt,
        ]
    table[x=L,y=AirComp_prod]
    {Data/Sim_L.dat};
    \addplot[ %smooth,
              %thin,
            color=airforceblue,
            mark=*,
            line width=0.9pt,
            mark size=1.5pt,
            ]
    table[x=L,y=ChannelComp_prod]
    {Data/Sim_L.dat};
    \addplot[ %smooth,
             %thin,
        color=cssgreen,
        mark=*,
        line width=0.9pt,
        mark size=1.5pt,
        ]
    table[x=L,y=ChannelCoding_prod]
    {Data/Sim_L.dat};
    % \legend{AirComp~\cite{hellstrom2023federated}, ChannelComp~\cite{saeed2023ChannelComp}, ChCompCode};
\end{axis}
\end{tikzpicture}
}
  \caption{Performance of ReChCompCode under different transmission slots in terms of NMSE averaged over $N_s=100$. Input values are $x_k=\{0,1,\ldots,7\}$ and the desired functions are $f = \sum_{k=1}^fx_k$ and $f=\prod_{k=1}^4x_k$  with $K=4$ nodes, respectively.
  }
\end{figure}
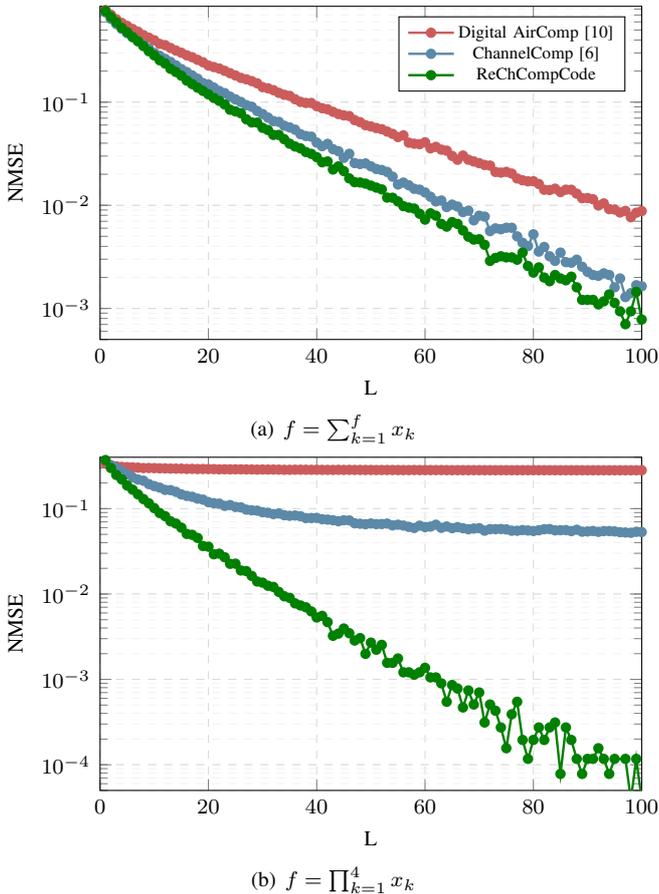

%====================================

For the first experiment, we consider a network of $K=4$ nodes for computing the product function $f=\prod_{k=1}^Kx_k$ and max function $f=\max_{k}x_k$, where $x_k\in \{0,1,\ldots,3\}$.  We make the comparison in terms of the \ac{NMSE} metric, $\text{\ac{NMSE}}:=\sum_{j=1}^{N_s}|f^{(i)}-\hat{f}_j^{(i)}|^2/(N_s|f^{(i)}|)$, defined as the sum of squared differences between desired function values $f^{(i)}$ and their estimated counterparts $\hat{f}_j^{(i)}$ divided by $N_s$ times the absolute value of $f^{(i)}$, where $N_s$ is the number of Monte Carlo trials. Since $L=2$ is enough to resolve the overlaps for given functions, we run the optimization problem in ~\eqref{eq:relaxed original problem} for $L=2$. Then, we repeat the pattern several times for transmission more than $L>2$.

Fig.~\ref{fig:four_node_product_SNR} and Fig.~\ref{fig:four_node_max_SNR} show the \ac{NMSE} of the function computation for the product and max functions, respectively over different \ac{SNR} levels, which is defined as $\text{\ac{SNR}}:=20\log(\|\bm{x}\|_2/\sigma_z)$. We observe a monotonic reduction in \ac{NMSE} for computing both functions. Especially within the \ac{SNR} range of  $5$~dB to $25$~dB, repetition yields rapid reduction results, which shows the effectiveness of the ReChCompCode method in concealing the channel noise.

\subsection{Comparison to AirComp and ChannelComp}

This subsection compares the  ReChCompCode method to \ac{AirComp} and ChannelComp. Specifically, we evaluate the ReChCompCode approach using two critical functions: summation and product functions denoted as $f=\sum_{k=1}^4x_k$, and $f=\prod_{k=1}^4x_k$, respectively. Here, the parameter $x_k$ takes values from the set $\{0, 1, \ldots, 7\}$, and the evaluation is carried out over a network comprising $K=4$ nodes. To ensure a fair comparison, we allocate the same power budget overall time slots, $L\times P_{\rm max}$, for all the methods.

Fig.~\ref{fig:four_node_sum_NMSE} and Fig.~\ref{fig:four_node_prod_NMSE} depict that \ac{NMSE} value for computing the summation and the product functions over different time slots $L$, respectively. ReChCompCode performs better than both \ac{AirComp} and ChannelComp in terms of \ac{NMSE}. In particular, ReChCompCode and ChannelComp show similar performance for computing the summation function. When computing the product function, since \ac{AirComp} approximates the product using the log function~\cite{csahin2023survey}, it cannot compute accurately even in high SNR scenarios. Additionally, due to the co-design of the channel code with constellation points, ReChCompCode achieves better performance than ChannelComp, which has already been proven to outperform OFDM~\cite{razavikia2023computing}. Notably, ReChCompCode reduces the computation error by approximately $30$~dB in \ac{NMSE} compared to ChannelComp for the product function with $100$ time slots.

\section{Conclusion}\label{sec:conclusion}

This paper introduced ReChCompCode, a channel code for repetition in digital over-the-air computation to provide a reliable communication protocol. Building upon the ChannelComp framework, we designed the channel coding scheme to reduce the computation error over \ac{MAC}. To this end, we proposed an optimization problem that jointly determines the encoding for digital modulation over multiple time slots. To manage the computational complexity of the proposed optimization problem, we developed an alternating minimization technique. Moreover, we evaluated the effectiveness of ReChCompCode through the numerical experiment by comparing it to existing state-of-the-art methods, such as digital \ac{AirComp} and ChannelComp. Notably, we observed approximately $30$~dB improvement in reducing the \ac{NMSE} of the computation error of the product function with $100$ time slots.

Looking forward, we plan to further study the convergence of the alternating minimization problem. Considering the challenge posed by its NP-hard complexity, we're also eager to explore simpler solutions. Notably, we anticipate demonstrating the substantial impact of this channel computing coding approach on applications such as federated edge learning. Additionally, we want to evolve from the current single antenna setup to a more comprehensive configuration with multiple inputs and outputs.

\bibliographystyle{IEEEtran}
\bibliography{Ref}

\appendix

\subsection{Proof of Proposition~\ref{pro:Lmin}}

We recall the constraints from the optimization problem in~\eqref{eq:relaxed original problem} as
\begin{align}
\sum\nolimits_{\ell=1}^L |(\bm{a}_i-\bm{a}_j)^\top(\bm{x} \odot \bm{c}_{\ell})|^2 \geq \Delta f_{i,j}, \label{eq:distance constraint}
\end{align} 
for all $(i,j) \in [M]^2$. Using H{\"o}lder Inequality~\cite{yang1991generalized}, we can bound the absolute value of the inner product as follows.
%----------------
\begin{align}
\nonumber
|(\bm{a}_i-\bm{a}_j)^\top(\bm{x} \odot \bm{c}_{\ell})|^2 & \leq \|\bm{a}_i-\bm{a}_j\|^2_1 \cdot \|\bm{x} \odot \bm{c}_{\ell}\|^2_\infty \\ \nonumber
& \leq \|\bm{a}_i-\bm{a}_j\|^2_1 \cdot \|\bm{x}\|^2_\infty \cdot \|\bm{c}_{\ell}\|^2_\infty \\
& \leq \|\bm{a}_i-\bm{a}_j\|^2_1 P_{\rm max}.
\end{align}
%----------------
Since the minimum distance between $\bm{a}_i$ and $\bm{a}_j$ is 1, by summing over all $L$ terms, we reach to
%----------------
\begin{align}
\nonumber
\sum_{\ell=1}^L |(\bm{a}_i-\bm{a}_j)^\top(\bm{x} \odot \bm{c}_{\ell})|^2 & \leq \sum_{\ell=1}^L\|\bm{a}_i-\bm{a}_j\|^2_1 \cdot \|\bm{x}\|^2_\infty \cdot \|\bm{c}_{\ell}\|^2_\infty \\
& \leq \frac{P_{\max}}{N} \cdot L. \label{eq:minimum distance}
\end{align}
%----------------
By substituting \eqref{eq:minimum distance} into \eqref{eq:distance constraint}, we obtain the lower bound of \(L\) as follows. 
%----------------
\begin{align}
L_{\min} \geq \frac{\max_{i,j} \Delta f_{i,j} \cdot N}{P_{\max}}.
\end{align}
%----------------
This concludes the proof. 
\end{document}